\documentclass[aps,pra,twocolumn,groupeaddress,longbibliography, preprintnumbers]{revtex4-1} 

\usepackage{amsmath}
\usepackage{amsfonts}
\usepackage{amssymb}
\usepackage{graphicx}
\usepackage{color}
\usepackage{accents}
\usepackage[colorlinks=true, citecolor=cyan]{hyperref}

\newcommand{\Tr}{\mbox{Tr}}

\newcommand{\ket}[1]{\left|#1\right\rangle}
\newcommand{\bra}[1]{\left\langle #1\right|}

\makeatletter 
\newsavebox{\@brx}
\newcommand{\llangle}[1][]{\savebox{\@brx}{\(\m@th{#1\langle}\)}%
  \mathopen{\copy\@brx\kern-0.5\wd\@brx\usebox{\@brx}}}
\newcommand{\rrangle}[1][]{\savebox{\@brx}{\(\m@th{#1\rangle}\)}%
  \mathclose{\copy\@brx\kern-0.5\wd\@brx\usebox{\@brx}}}
\makeatother

\newlength{\dhatheight} 

\newcommand{\qed}{\nobreak \ifvmode \relax \else
      \ifdim\lastskip<1.5em \hskip-\lastskip
      \hskip1.5em plus0em minus0.5em \fi \nobreak
      \vrule height0.75em width0.5em depth0.25em\fi}

\begin{document}

\title{Robust quantum sensing in strongly interacting systems with many-body scars}
\author{Shane Dooley}
\email[]{dooleysh@gmail.com}
\affiliation{Dublin Institute for Advanced Studies, School of Theoretical Physics, 10 Burlington Rd, Dublin, Ireland}
\date{\today}
\preprint{DIAS-STP-21-02}

\begin{abstract}
  In most quantum sensing schemes, interactions between the constituent particles of the sensor are expected to lead to thermalisation and degraded sensitivity. However, recent theoretical and experimental work has shown that the phenomenon of quantum many-body scarring can slow down, or even prevent thermalisation. We show that scarring can be exploited for quantum sensing that is robust against certain strong interactions. In the ideal case of perfect scars with harmonic energy gaps, the optimal sensing time can diverge despite the strong interactions. We demonstrate the idea with two examples: a spin-1 model with Dzyaloshinskii-Moriya interaction, and a spin-1/2 mixed-field Ising model. We also briefly discuss some non-ideal perturbations, and the addition of periodic controls to suppress their effect on sensing.
\end{abstract}


\maketitle

\section{Introduction}

The estimation of physical quantities is an important task in many branches of science and technology. In recent decades, the field of quantum metrology and sensing has emerged, in which the objective is to exploit quantum coherence or entanglement to give enhanced sensitivity in such parameter estimation tasks \cite{Gio-11, Deg-17, Pez-18}. Proposed applications include magnetometry \cite{Bud-07, Tay-08, Tan-15}, electrometry \cite{Dol-11, Fac-16}, quantum clocks \cite{Lud-15} and, perhaps most famously, gravitational wave detection \cite{Cav-81, Aas-13}.






One of the most widely used approaches to quantum sensing is Ramsey interferometry, or one of its variants. After preparing the probe system in some quantum superposition state, it is allowed to interact with the parameter-of-interest, followed by a measurement of the probe to extract information about the parameter \cite{Yur-86b, Lee-02, Gio-11, Deg-17, Pez-18}. Precise sensing then usually relies on maintaining quantum coherence in the probe system for long times. However, in many realistic sensing devices, other unwanted interactions are expected to degrade the achievable sensitivity by limiting the useful coherence time. Even if the probe is completely isolated from any external environment, internal interactions between the consituent particles of the probe can lead to decoherence and thermalisation. Moreover, a high density of probe particles will likely result in stronger interactions, and more rapid decoherence and thermalisation. Several schemes to overcome this limitation have been proposed \cite{Doo-18a, Cho-17, Rag-18}, or implemented experimentally \cite{Zho-20}.

Recently, a new mechanism was discovered in which thermalisation can be slowed down, or even completely avoided, despite strong interactions in a non-integrable many-body system. This mechanism -- dubbed quantum many-body scarring -- was shown to be responsible for long-lived oscillations in an experiment on a chain of interacting Rydberg atoms \cite{Ber-17, Tur-18a}. Since the long-lived oscillations are associated with long coherence times, this is suggestive of a possible advantage in quantum sensing \cite{Ser-20}. Despite intensive work on scars and their properties \cite{Mou-18a,Mou-18b,Ok-19,Bul-19,Ho-19,Mar-20,Shi-20,Mou-20,Mic-20,Iad-20,Bul-20,Kun-20,McC-20,Ban-20, Doo-20b}, this possibility has not yet been explored.


In this paper, we examine the potential for robust quantum sensing via quantum scarring. We begin in section \ref{sec:general} with a brief discussion of quantum sensing, and its connection with many-body scars. Then, through two examples, we demonstrate the connection more concretely. First, in section \ref{sec:spin_1}, a spin-1 model with a Dzyaloshinskii-Moriya interaction (DMI). We find that despite the non-integrability of the model, robust quantum sensing is possible. This is associated with a diverging coherence time caused by of a set of scars with perfectly harmonic energy gaps. Then, in section \ref{sec:MFIM}, we consider a mixed-field Ising model. Usually, strong Ising interactions between the spins are expected to lead to fast decoherence and thermalisation. Counterintuitively, in this example the stronger couplings can extend the coherence time and enhance the quantum sensing. We show that this is due to the emergence of a set of quantum many-body scars in the ``PXP'' limit of the mixed-field Ising model. Finally, in section \ref{sec:noise} we briefly discuss some non-ideal perturbations, and the possibility of suppressing their effect on sensing by periodic driving.

\section{Quantum sensing, ETH, and many-body scars}\label{sec:general}



Consider an $N$-particle probe system, whose purpose is to estimate a parameter $\omega$ that appears in its Hamiltonian $\hat{H} = \omega \sum_{n=0}^{N-1}\hat{h}_n + \hat{H}_\text{int}$. Here, $\hat{h}_n$ is a local operator for the $n$'th particle and $\hat{H}_\text{int}$ generates interactions between the particles. Typically, the estimation scheme involves initialising the probe in some easily prepared state $\ket{\psi(t=0)}$, and extracting information about the parameter $\omega$ from a measurement of the time-evolved state $\ket{\psi(t)} = e^{-it\hat{H}}\ket{\psi(0)}$. For small probe systems (e.g., a single qubit) arbitrary measurements of the final state may be possible. For larger systems, however, the measurement may be restricted to observables $\hat{\mathcal{O}} \in \mathcal{M}$ from some set of experimentally accessible measurements $\mathcal{M}$. This set will depend on the details of the physical implementation, but is often a subset of local operators. To accumulate statistics about the parameter-of-interest, the prepare-evolve-measure sequence is repeated during a total available time $T$, giving $T/t$ independent repetitions of the measurement (where the time needed for state preparation and readout is assumed to be negligible \cite{Doo-16b, Hay-18}). Employing method-of-moments estimation, the final error can be calculated by the propagation-of-error formula \cite{Win-94, Pez-18}: \begin{equation} \delta\omega = \min_{\hat{\mathcal{O}} \in \mathcal{M}} \frac{ \Delta \hat{\mathcal{O}}(t)}{| \partial_{\omega} \langle \hat{\mathcal{O}} (t) \rangle | \sqrt{T/t}} , \label{eq:error_formula} \end{equation} where the numerator is the uncertainty $\Delta\hat{\mathcal{O}} \equiv \sqrt{\langle \hat{\mathcal{O}}^2\rangle - \langle\hat{\mathcal{O}} \rangle^2}$ of the measured observable $\langle\hat{\mathcal{O}}\rangle$ and the factor $|\partial_\omega \langle\hat{\mathcal{O}}\rangle|$ in the denominator quantifies its response to small changes in the parameter $\omega$.


In the setup described above, the dynamics are unitary and the time-evolved state $\ket{\psi(t)}$ is always pure. However, for a non-integrable many-body system the interactions $\hat{H}_\text{int}$ between particles can lead to decoherence and thermalisation with respect to the expectation values $\langle\hat{\mathcal{O}}\rangle$. This is because the evolution generates entanglement in the system that makes information about the initial conditions inaccessible to the experimental observables $\hat{\mathcal{O}} \in \mathcal{M}$.

Instead of waiting for the probe to thermalise, it is usually better to measure the system before it thermalises, at some optimal sensing time $t = t_*$ giving the optimised error $\delta\omega_* = \min_t \delta\omega$. For separable initial states $\ket{\psi(0)}$ the optimised error typically has the form $\delta\omega_* \propto 1/\sqrt{N t_* T}$, with fast decoherence and thermalisation corresponding to a short $t_*$, and poor quantum sensor performance \cite{Deg-17}. To estimate $\omega$ precisely, one should try to engineer a probe system that has both a large coherence time $t_*$ and a large number of particles $N$. One approach is to design the probe Hamiltonian so that the interacting part is suppressed as much as possible, $\hat{H}_\text{int} \approx 0$. This gives a long coherence time $t_*$, but usually means that the particle density must be low, to suppress the interactions. For high density quantum sensing, with long coherence times $t_*$, we should look for mechanisms for avoiding decoherence and thermalisation, even in the presence of strong interactions.

The process of thermalisation in closed quantum systems is often framed in terms of the eigenstate thermalisation hypothesis (ETH) \cite{Deu-91, Sre-94, Rig-08}. Let $\hat{H} = \sum_j E_j \ket{E_j}\bra{E_j}$ be the spectral decomposition of the Hamiltonian. An eigenstate $\ket{E_j}$ is said to be thermal if, for all $\hat{\mathcal{O}} \in \mathcal{M}$, we have $\bra{E_j}\hat{\mathcal{O}}\ket{E_j} \approx \Tr[\hat{\mathcal{O}}\hat{\rho}_\text{th}]$, where $\hat{\rho}_\text{th}$ is the thermal state at the temperature corresponding to the energy $E_j$. One can show that if \emph{all} eigenstates around a given finite energy density are thermal, the observables $\hat{\mathcal{O}}\in\mathcal{M}$ will thermalise for any initial state at that energy density \cite{DAl-16}. Conversely, the system can fail to thermalise if there are some non-thermal eigenstates that have a large overlap with the initial state \cite{Bir-10}. Recently, such ETH-violating systems were discovered where most, but not all, Hamiltonian eigenstates are thermal \cite{Shi-17}. The non-thermal eigenstates were dubbed quantum many-body scars (QMBS), and were found to be responsible for long-lived coherence in an experiment with a chain of Rydberg atoms \cite{Ber-17, Tur-18a, Ser-20}. 


Based on the foregoing discussion, some necessary conditions for robust quantum sensing via many-body scars are:

\begin{enumerate}

\item A Hamiltonian with a set of QMBS.

\item An easily prepared initial state, with a large component in the QMBS subspace.

\item Dynamics in the QMBS subspace that are sensitive to the Hamiltonian parameter to be estimated.

\item An experimentally accessible observable $\hat{\mathcal{O}} \in \mathcal{M}$ that can extract the parameter information from the time-evolved state.

\end{enumerate}

In sections \ref{sec:spin_1} and \ref{sec:MFIM} we present two example models in which these criteria are satisfied, giving robust quantum sensing despite strong interactions in the many-body probe system. Both examples suggest that in the ideal case (perfect scars with harmonic energy gaps, and full overlap with the initial state) the optimal sensing time $t_*$ diverges. 

\section{Example: spin-1 DMI model}\label{sec:spin_1}







\subsection{Quantum sensing}\label{sec:spin_1_sensing}


As our first example, we consider a sensing scheme where the goal is the estimation of a magnetic field $\omega$ acting on a system of $N$ interacting spin-1 particles via the Hamiltonian: \begin{equation} \hat{H}_0 = \frac{\omega}{2} \sum_{n=0}^{N-1} \hat{S}_n^z , \end{equation} where $\hat{S}^z = \sum_{m\in\{-1,0,1\}}m\ket{m}\bra{m}$. However, the particles may also interact with each other via the Hamiltonian \begin{equation} \hat{H}_\text{int} = \sum_{n,n'} \lambda_{n,n'} (e^{i\phi} \hat{S}_n^+ \hat{S}_{n'}^{-} + \text{h.c.}) , \label{eq:H_int} \end{equation} where $\hat{S}^\pm = \sum_{m\in\{-1,0,1\}} (1 - \delta_{m, \pm 1}) \ket{m \pm 1}\bra{m}$ are the spin raising and lowering operators, giving a total Hamiltonian $\hat{H} = \hat{H}_0 + \hat{H}_\text{int}$. The spin-spin coupling parameters $\lambda_{n,n'}$ are assumed to be real but are otherwise arbitrary and can, for example, be any real function of the positions of the interacting spins in some $d$-dimensional space. The interaction Hamiltonian can be rewritten as $\hat{H}_\text{int} = \sum_{n,n'}\lambda_{n,n'} [ \cos\phi (\hat{S}_n^x\hat{S}_{n'}^x + \hat{S}_n^y\hat{S}_{n'}^y) + \sin\phi (\hat{S}_n^x\hat{S}_{n'}^y - \hat{S}_n^y\hat{S}_{n'}^x)]$, showing that the phase $\phi$ rotates between an XX-interaction for $\phi \in \{ 0, \pm\pi \}$, and a Dzyaloshinskii-Moriya interaction (DMI) for $\phi = \pm\pi/2$.



After preparing the spins in the initial product state: \begin{eqnarray} \ket{\psi(t=0)} = \ket{\boldsymbol{+}} \equiv \bigotimes_{n=0}^{N-1} \frac{\ket{m_n = +1} + \ket{m_n = -1}}{\sqrt{2}} , \label{eq:psi_plus} \end{eqnarray} we suppose that the system is allowed to evolve by the Hamiltonian $\hat{H}$ for a sensing time $t$, followed by a measurement of the local observable $\hat{\mathcal{O}}_\theta = e^{-i\theta}\sum_n (\hat{S}_n^+)^2 + \text{h.c.}$, where $\theta$ can be tuned for the optimal sensing performance. This prepare-evolve-measure sequence is repeated during a total available time $T$, giving $T/t$ independent repetitions of the measurement.  The error in the estimate of $\omega$ is calculated by the propagation-of-error formula given in Eq. \ref{eq:error_formula}. We note that the initial product state in Eq. \ref{eq:psi_plus} is usually relatively easy to prepare in practice. For example, after cooling to the ground state $\otimes_n \ket{m_n = -1}$ of $\hat{H}$ in a strong magnetic field, a collective rotation of all spins can generate the desired state.






\begin{figure} 
  \includegraphics[width=\columnwidth]{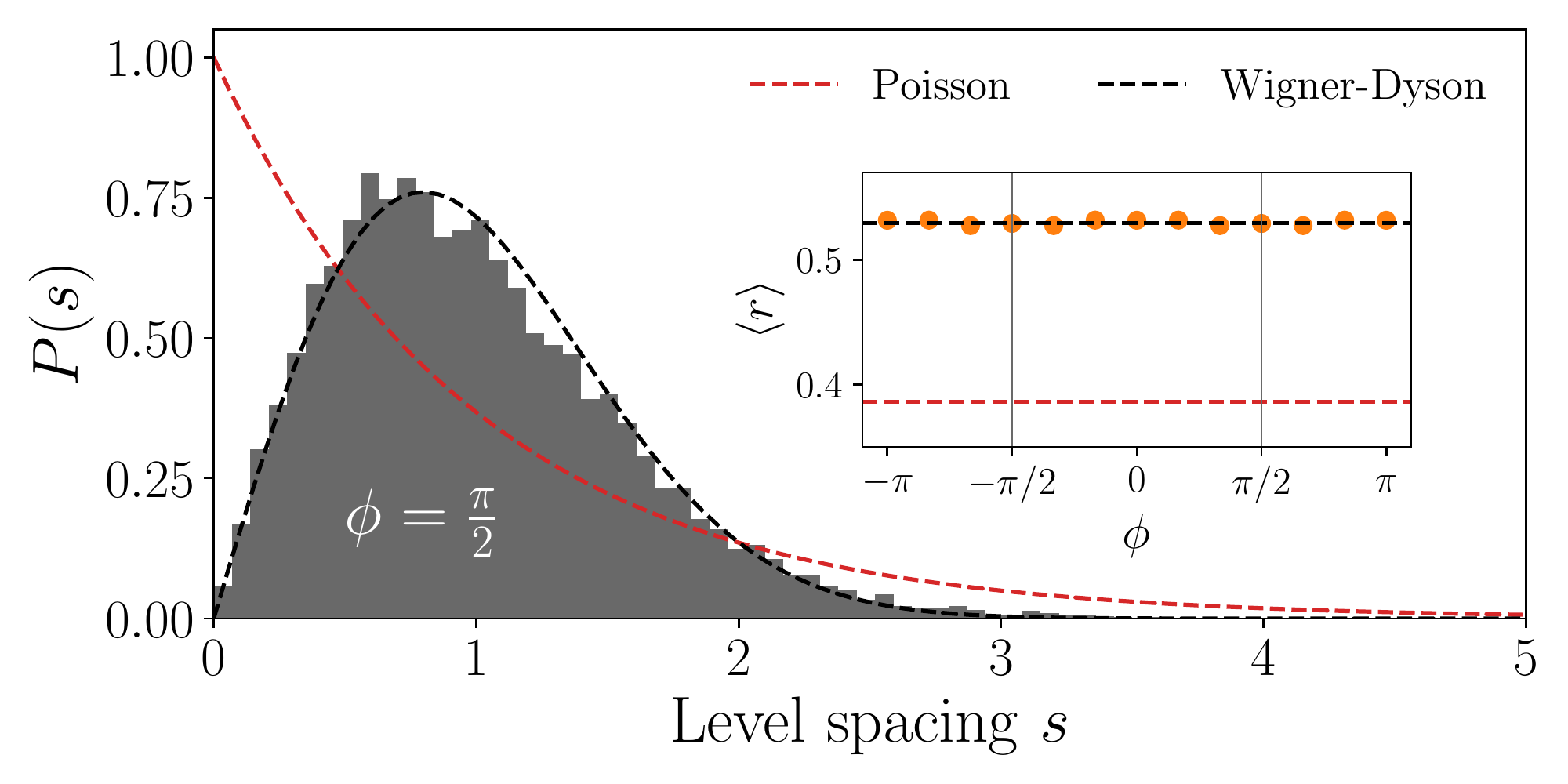}
  \caption{Here, we demonstrate the non-integrability of $\hat{H}$ for all values of $\phi$, in a $d=1$ dimensional example with $\lambda_{n,n'} = \lambda/(n-n')^2$ and periodic boundary conditions. The Wigner-Dyson distribution $P(s)$ for the energy spacings $s_i = E_{i+1} - E_i$, and the associated $r$-value of $\langle r \rangle \approx 0.53$ are signatures of non-integrability.   All quantities are calculated in the $k = 0$ momentum, and $\sum_n \hat{S}_n^z = 1$ magnetization symmetry sectors. For $\phi \in \{ 0, \pm\pi\}$ we must also restrict to the even reflection parity sector. [Other parameters: $N = 14$, $\omega = 1$, $\lambda = 0.5$.]} 
  \label{fig:level_statistics}
\end{figure}


As a benchmark, we first calculate the error in the case when there are no interactions between the spins, i.e., when $\lambda_{n,n'} = 0$ for all $n$, $n'$. Then, the time-evolved state is: \begin{eqnarray} \ket{\psi(t)} = \bigotimes_{n=0}^{N-1}\frac{e^{-it\omega/2}\ket{m_n=+1} + e^{it\omega/2}\ket{m_n=-1}}{\sqrt{2}} . \label{eq:psi_t_non_int} \end{eqnarray} We see that the dynamics are periodic, with period $\tau = 2\pi/\omega$, and the spins remain in a product state throughout the evolution. The error, easily calculated by the formula in Eq. \ref{eq:error_formula}, is then equal to the standard quantum limit, given by $\delta\omega = \delta\omega_\text{SQL} \equiv 1/\sqrt{NtT}$.

However, the non-interacting model is a very special case, for which the Hamiltonian $\hat{H}$ is integrable. Generally, for $\hat{H}_\text{int} \neq 0$, $\hat{H}$ is non-integrable. The interactions between the spins might therefore be expected to result in a degradation of the sensing performance. Indeed, when interactions are present the error typically does not decrease monotonically with the sensing time $t$. Rather, it decreases to a minimum value $\delta\omega_* \equiv \min_t \delta\omega$ at an optimal sensing time $t_*$.

\begin{figure}
  \includegraphics[width=\columnwidth]{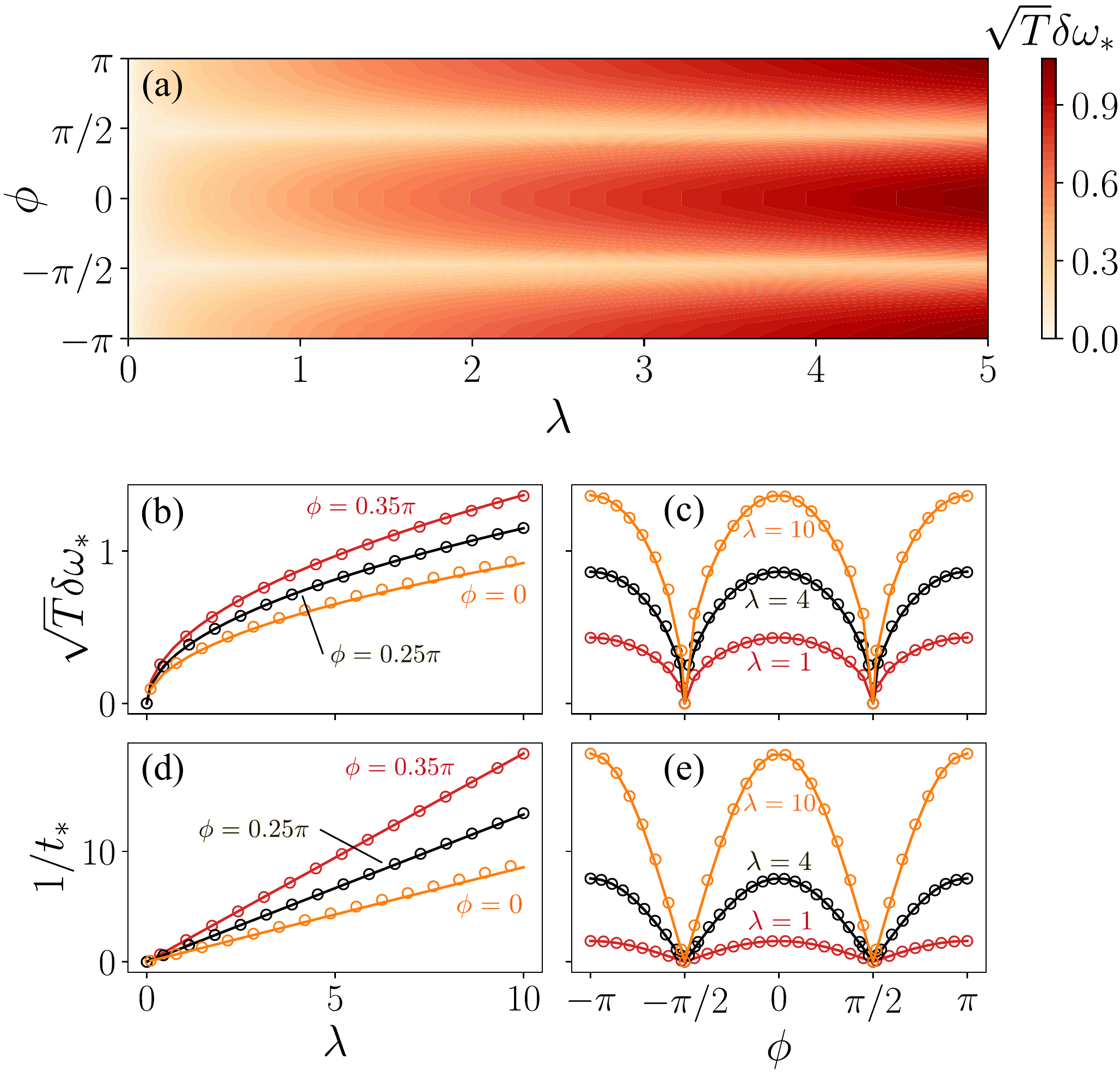}
  \caption{(a) The optimised error $\delta\omega_*$, calculated numerically for the $d=1$ example with $\lambda_{n.n'} = \lambda/(n-n')^2$, periodic boundary conditions, and $N=10$. The plots in (b) and (c) are cross sections of (a) (although for $N=12$). (d), (e) show the corresponding optimal sensing times $t_*$. The numerical data (the circles) are closely matched by the equations $\delta\omega_* = 1.09/\sqrt{Nt_* T}$ and $t_* = 0.53/|\lambda \cos\phi|$ (the solid lines). [All plotted for $\omega = 1$, although the error $\delta\omega$ is independent of $\omega$; see Appendix \ref{app:omega_independence}.]}
  \label{fig:opt_estimation_error} 
\end{figure}

We illustrate these points for an example in $d=1$ dimensions, with an interaction $\lambda_{n,n'} = \lambda/(n-n')^2$ and assuming periodic boundary conditions. We show in Fig. \ref{fig:level_statistics}, by calculating energy level spacing statistics, that this Hamiltonian is non-integrable for any choice of the phase $\phi$. In Fig. \ref{fig:opt_estimation_error} we plot the optimised error $\delta\omega_*$ and the optimal sensing time $t_*$, as a function of the overall coupling strength $\lambda$ and the phase $\phi$. We see that the equations: \begin{equation} \delta\omega_* = \frac{1.09}{\sqrt{Nt_{*}T}} , \quad t_* = \frac{0.53}{|\lambda\cos\phi|} , \label{eq:error_fit} \end{equation} fit the numerical data very well. It is clear that the optimal sensing time is long, and the error is low, when the interaction strength $\lambda$ is small. This is not surprising since it is close to the case of non-interacting spins precessing in the field $\omega$. However, the other notable feature of Fig. \ref{fig:opt_estimation_error} and Eq. \ref{eq:error_fit} is the diverging optimal sensing time $t_* \to \infty$ when $\phi = \pm \pi/2$, even for strong interactions $\lambda$. This may be surprising, considering the non-integrability of the Hamiltonian. We now show that the diverging $t_*$, and the associated low error for $\phi \approx \pm \pi/2$, is due to quantum many-body scarring, and is a feature of the model not only for our $d=1$ dimensional example, but for any choice of the interaction strengths $\lambda_{n,n'}$, in any spatial dimension.

\subsection{Quantum many-body scars in the spin-1 DMI model}\label{sec:spin_1_QMBS}

Before proceeding, it is convenient to introduce the spin-1/2 operators that are obtained by restricting to the $\ket{m_n = \pm 1}$ local basis states of each spin-1 particle. These operators are $\hat{\sigma}_n^\pm \equiv (\hat{S}_n^\pm)^2 = \ket{m_n = \pm 1}\bra{m_n = \mp 1}$, $\hat{\sigma}_n^z = [\hat{\sigma}_n^+, \hat{\sigma}_n^-] = \hat{S}_n^z$, with the associated spin-1/2 collective operators $\hat{J}^\pm \equiv \sum_{n=0}^{N-1}\hat{\sigma}_n^\pm$ and $\hat{J}^z \equiv \frac{1}{2}[\hat{J}^+, \hat{J}^-] = \frac{1}{2}\sum_{n=0}^{N-1}\hat{\sigma}_n^z$. The symmetric Dicke states $\{\ket{\Psi(s)}\}_{s=0}^N$ are defined as simultaneous eigenstates of $\hat{J}^z$ and $\hat{J}^2 \equiv (\hat{J}^x)^2 + (\hat{J}^y)^2 + (\hat{J}^z)^2$, and can be written as \cite{Dic-54}: \begin{equation} \ket{\Psi(s)} \equiv \mathcal{N}_s (\hat{J}^+)^s \ket{\Downarrow} , \quad s \in \{ 0,1, \hdots, N \} , \label{eq:dicke_states} \end{equation} where $\mathcal{N}_s = \frac{1}{s!}\binom{N}{s}^{-1/2}$ is a normalisation factor and $\ket{\Downarrow} \equiv \bigotimes_{n=0}^{N-1}\ket{m_n = -1}$. 


It was recently shown \cite{Mar-20b} that the symmetric Dicke states in Eq. \ref{eq:dicke_states} are scar states of the Hamiltonian $\hat{H}$ when $\phi = \pm\pi/2$. In other words, they are ETH-violating eigenstates of $\hat{H}(\phi=\pm\pi/2)$, with a sub-volume law growth of entanglement entropy. We already know that they are eigenstates of the non-interacting part of the Hamiltonian $\hat{H}_0\ket{\Psi(s)} = \omega(s-N/2)\ket{\Psi(s)}$, since this is one of the defining properties of Dicke states. Writing $\hat{h}_{n,n'}(\phi) \equiv e^{i\phi}\hat{S}_n^+\hat{S}_{n'}^- + e^{-i\phi}\hat{S}_n^-\hat{S}_{n'}^+$, in Appendix \ref{app:Dicke_scars} we also show that $\hat{h}_{n,n'}(\pm\pi/2) \ket{\Psi(s)} = 0$ for all $n$, $n'$, implying that $\hat{H}_\text{int}\ket{\Psi(s)} = 0$. Therefore, the Dicke states are eigenstates of the full Hamiltonian $\hat{H} = \hat{H}_0 + \hat{H}_\text{int}$, with the eigenvalue equation: \begin{equation} \hat{H}(\phi=\pm\pi/2) \ket{\Psi(s)} = \omega(s-N/2)\ket{\Psi(s)} . \label{eq:dicke_scar} \end{equation}

Since our initial product state, given in Eq. \ref{eq:psi_plus}, can be rewritten in terms of the Dicke states $\ket{\boldsymbol{+}} = 2^{-N/2} \sum_{s=0}^{N}\binom{N}{s}^{1/2} \ket{\Psi(s)}$, we can use Eq. \ref{eq:dicke_scar} to write the time-evolved state: \begin{eqnarray} \ket{\psi(t)} &=& 2^{-N/2} \sum_{s=0}^{N}\binom{N}{s}^{1/2} e^{-it \omega (s-N/2)}\ket{\Psi(s)} \\ &=& \bigotimes_{n=0}^{N-1}\frac{e^{-it\omega/2}\ket{m_n=+1} + e^{it\omega/2}\ket{m_n=-1}}{\sqrt{2}} , \qquad \label{eq:psi_t_scar} \end{eqnarray} showing that the evolution takes place entirely within the Dicke scar subspace. The product state Eq. \ref{eq:psi_t_scar} is identical to the non-interacting time-evolved state given in Eq. \ref{eq:psi_t_non_int}. The error is therefore the same, $\delta\omega = \delta\omega_\text{SQL} = 1/\sqrt{NtT}$, despite the non-zero interactions and the non-integrability of the Hamiltonian. As the error is always decreasing with time it has no minimum value, i.e., the optimal sensing time diverges $t_* \to \infty$ as indicated by the numerical results in Fig. \ref{fig:opt_estimation_error}.


One might wonder if the emergence of scars (and the associated robustness in the sensing) relies on special symmetries of the spin-1 Hamiltonian at $\phi = \pm\pi/2$. Indeed, a feature of the Hamiltonian at $\phi = \pm\pi/2$ is that it has the ``particle-hole'' symmetry $\{ \hat{\Pi},\hat{H} \} = 0$, where $\hat{\Pi} = \bigotimes_n [\ket{0_n}\bra{0_n} + \ket{1_n}\bra{-1_n} + \ket{-1_n}\bra{1_n}]$. However, as was noted in Refs. \cite{Sch-19, Mar-20b}, this symmetry can be broken by including a term $\hat{H}_D = D\sum_n (\hat{S}_n^z)^2$ in the Hamiltonian, without disturbing the Dicke scar states $\{ \ket{\Psi(s)} \}$. The associated sensing performance therefore remains at the standard quantum limit when $\phi=\pm\pi/2$, despite the addition of $\hat{H}_D$ to the Hamiltonian.

\begin{figure}
  \includegraphics[width=\columnwidth]{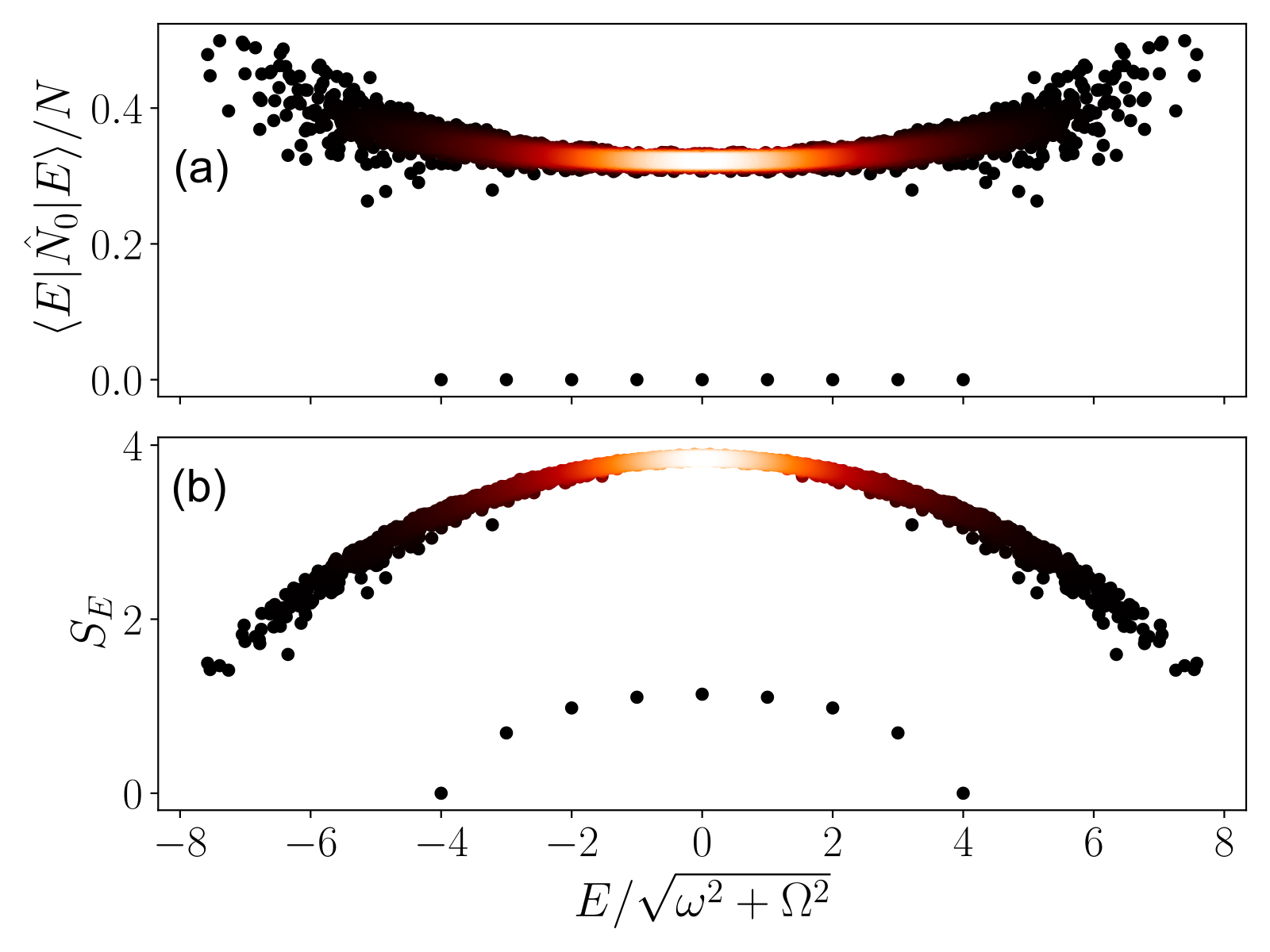}
  \caption{(a) Expectation values of the observable $\hat{N}_0$, and (b) the half-chain entanglement entropy $S_E$, clearly show the violation of the ETH for the Dicke scar states (the band of isolated states separated from the bulk, with low entropy and $\langle \hat{N}_0 \rangle = 0$). The color scale corresponds to the density of data points, with high-density regions in yellow and low-density regions in black. Parameters: $N = 8$, $\omega = 1.0$, $\Omega = 0.3$, $\eta=D=0$, $\phi = \pi/2$, $\lambda_{n,n'} = c_{n,n'}/(n-n')^2$ where $c_{n,n'}$ is chosen at random from the interval $[0.5,1.0]$, to break translation invariance.}
  \label{fig:eigvec_exp_vals_and_ent_entropy} 
\end{figure}

The spin-1 Hamiltonian $\hat{H}$ also has a $U(1)$ symmetry associated with the conserved total magnetization $[\hat{H}, \sum_n \hat{S}_n^z] = 0$. However, this symmetry can be broken by adding the local term $\hat{H}_\Omega = \frac{\Omega}{2} \sum_n[ e^{i\eta} (\hat{S}_n^+)^2 + e^{-i\eta} (\hat{S}_n^-)^2]$ to the Hamiltonian. Then the non-interacting part of the Hamiltonian can be written as $\hat{H}_0 + \hat{H}_\Omega = \sqrt{\omega^2 + \Omega^2} \; \hat{U}\hat{H}_0\hat{U}^\dagger$, where $\hat{U} \equiv \exp[\frac{1}{2}e^{-i\eta}\arctan\left(\frac{\Omega}{\omega}\right)\sum_n(\hat{S}_n^-)^2 + \text{h.c.}] $. Despite the broken magnetization symmetry, the full spin-1 Hamiltonian has the rotated set of Dicke scar states $\{\hat{U}\ket{\Psi(s)}\}$ (this is shown in Appendix \ref{app:Dicke_scars}). If we consider a sensing scheme to estimate the Hamiltonian parameter $\sqrt{\omega^2 + \Omega^2}$, with the initial state $\hat{U}\ket{\boldsymbol{+}}$ and the measurement observable $\hat{U}\mathcal{O}_\theta\hat{U}^\dagger$, then the error is again at the standard quantum limit. It appears that the only symmetry that cannot be broken in the Hamiltonian $\hat{H}(\phi=\pm\pi/2)$ without destroying the scars, is the number parity symmetry $[\hat{H},\hat{P}_0]=0$, where $\hat{P}_0 \equiv (-1)^{\hat{N}_0}$ and $\hat{N}_0 \equiv \sum_n \ket{m_n=0}\bra{m_n=0}$.

To verify that the ETH is violated by the Dicke states we consider the expectation values $\bra{E}\hat{N}_0\ket{E}/N$ for eigenstates $\ket{E}$ of the Hamiltonian. For convenience, we suppose that the Hamiltonian has no symmetry except for the conservation of number parity $[\hat{H},\hat{P}_0]=0$. For such an example we expect an infinite temperature thermal eigenstate to have $\bra{E}\hat{N}_0\ket{E}/N \approx 1/3$ (i.e., $1/3$ probability of being in each of the three local basis states $m\in\{-1,0,1\}$). However, the Dicke states all have $\bra{\Psi(s)} \hat{N}_0 \ket{\Psi(s)} = 0$. This ETH-violation is shown numerically for an example in Fig. \ref{fig:eigvec_exp_vals_and_ent_entropy}(a). The half-chain entanglement entropy for each eigenstate in the same example is shown in Fig. \ref{fig:eigvec_exp_vals_and_ent_entropy}(b). We note that it is possible to analytically calculate the entanglement entropy of any Dicke state $\ket{\Psi(s)}$ for any bipartition of the $N$ spins \cite{Mor-18b}. The scar state with the largest entanglement entropy is the $s=N/2$ Dicke state. For a bipartition into two equal-size clusters of spins, its entanglement entropy tends to $S \to \frac{1}{2}\log(N/2)$ as $N\to\infty$. All of the Dicke states therefore have a sub-volume law growth of entanglement entropy \cite{Sch-19}.

\section{Example: spin-1/2 MFI model}\label{sec:MFIM}

\subsection{Quantum sensing}

The robust quantum sensing via many-body scars is not specific to the spin-1 DMI model. In this example we consider a sensing protocol in which the goal is the estimation of the transverse field parameter $\omega$ in a chain of $N$ spin-$1/2$ particles with the mixed-field Ising (MFI) Hamiltonian:  \begin{equation} \hat{H}_\text{MFI} = \frac{\omega}{2} \sum_{n=0}^{N-1} \hat{\sigma}_n^x + \frac{\Omega}{2} \sum_{n=0}^{N-1} \hat{\sigma}_n^z + \frac{\lambda}{4} \sum_{n=0}^{N-1} \hat{\sigma}_n^{z}\hat{\sigma}_{n+1}^z . \label{eq:H_MFIM} \end{equation} Here $\hat{\sigma}_n^\mu$, $\mu \in \{x,y,z \}$, are the spin-1/2 Pauli operators at site $n$ on the chain and we assume the periodic boundary conditions $\hat{\sigma}_{n+N}^\mu \equiv \hat{\sigma}_{n}^\mu$.

\begin{figure}
  \includegraphics[width=\columnwidth]{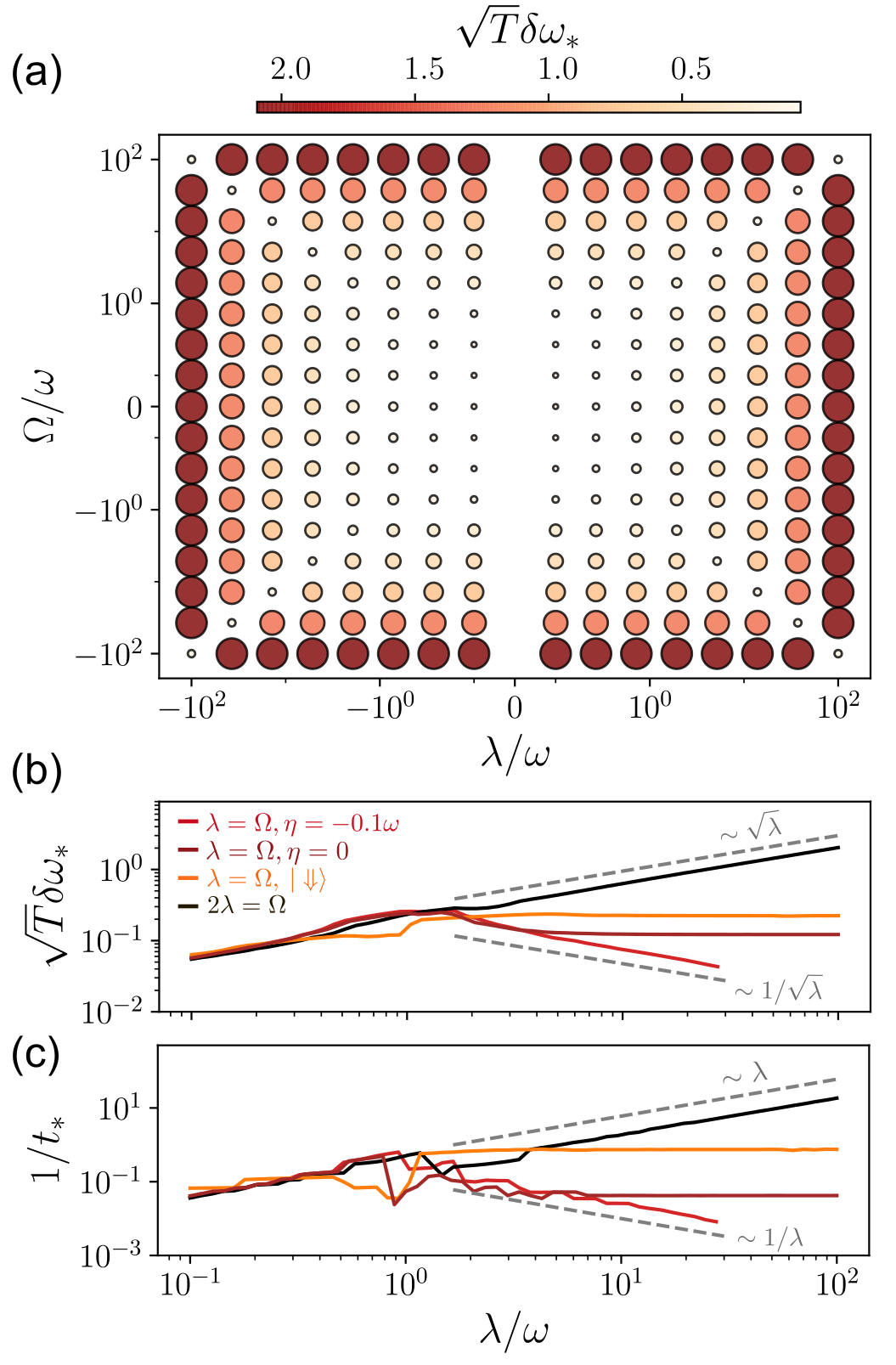}
  \caption{(a) The optimised error $\delta\omega_*$ is represented by both the size and color of the marker. The error is low when $\{ |\Omega|, |\lambda| \} \ll \omega$, since this is close to the ideal case of non-interacting spins. More interestingly, the error is also low when $|\Omega| = |\lambda| \gg \omega$, corresponding to the regime of quantum many-body scarring. (b) As quantum scars emerge the error decreases as $\delta\omega_* \sim |\lambda|^{-1/2}$ for the inital N\'{e}el state $\ket{\mathbb{Z}_2}$ (red lines). This is particularly clear when the scars are enhanced by the perturbation $\eta = -0.1\omega$ to the Hamiltonian (see Eq. \ref{eq:delta_H}). In contrast, outside the regime of QMBS the error increases as $\delta\omega_* \sim |\lambda|^{1/2}$ (black line). For the initially polarised state $\ket{\Downarrow} \equiv \ket{\downarrow\downarrow\downarrow...}$ the error plateaus, but does not decrease as $\lambda$ increases (orange line). (c) As perfect scars emerge the optimal sensing time diverges as $t_* \sim |\lambda|$. [Parameters (unless otherwise stated in legend): $N = 16$; $\ket{\psi(0)} = \ket{\mathbb{Z}_2}$, $\eta = 0$, $\omega = 1$.]}
  \label{fig:main}
\end{figure} 

After preparing the spins in the initial N\'{e}el state $\ket{\psi(0)} = \ket{\mathbb{Z}_2} \equiv \ket{\uparrow\downarrow\uparrow\downarrow\hdots}$, the spin system evolves by the Hamiltonian $\hat{H}_\text{MFI}$ for a sensing time $t$, followed by the measurement of an observable $\hat{\mathcal{O}} \in \mathcal{M}$, where $\mathcal{M}$ is some set of experimentally accessible observables. The measurement is repeated $T/t$ times, resulting in the estimation error given in Eq. \ref{eq:error_formula}. For the purposes of this example it is sufficient to consider the space of accessible measurement observables $\mathcal{M}$ to be spanned by the basis set $\{ \hat{J}_\text{odd}^\mu , \hat{J}_\text{even}^\mu \}_{\mu\in\{x,y,z\}}$, where $\hat{J}_\text{odd/even}^\mu \equiv \frac{1}{2}\sum_{n \text{ odd/even}} \hat{\sigma}_n^\mu $. A measurement observable is thus of the form $\hat{\mathcal{O}} = \sum_{\mu\in\{x,y,z\}} (c_\text{odd}^\mu \hat{J}_\text{odd}^\mu + c_\text{even}^\mu \hat{J}_\text{even}^\mu)$ for $c_\text{odd/even}^\mu \in \mathbb{R}$. We note that, in practice, such a measurement can be implemented if it is possible to perform arbitrary collective rotations of the odd/even spins separately, just prior to the measurement of a single collective observable, e.g., the total magnetization $\hat{J}^z = \hat{J}_\text{odd}^z + \hat{J}_\text{even}^z$. Such operations do not require single-site addressability of the spins. Also, the initial N\'{e}el state $\ket{\mathbb{Z}_2}$ is often easily prepared, for example by cooling to the ground state of a Hamiltonian with a strong anti-ferromagnetic interaction, or a nearest-neighbour blockade (as in the experiments of Ref. \cite{Ber-17}).


In the case of non-interacting spins without a longitudinal field ($\Omega = \lambda = 0$) the time-evolved state is a product state. The error is easily calculated and is given by the standard quantum limit $\delta\omega_\text{SQL} = 1/\sqrt{NtT}$. For $\lambda \neq 0$ the error no longer decreases monotonically in time, and so we focus our attention on the minimum error $\delta\omega_* = \min_t \delta\omega$. If, in addition, $\Omega \neq 0$ the Hamiltonian is non-integrable, and we resort to numerical simulation to obtain our results. These results are summarised in Fig. \ref{fig:main}(a) where we plot the optimised error $\delta\omega_* = \min_t \delta\omega$ for various values of the longitudinal field $\Omega$ and the Ising coupling $\lambda$. We see that the error $\delta\omega_*$ is small when $\{|\Omega|, |\lambda|\} \ll |\omega|$. This is not surprising, since it is close to the case of non-interacting spins precessing in the transverse field $\omega$. As the interaction strength $\lambda$ increases the approximation to non-interacting spins begins to break down, which we expect to degrade the error. Fig. \ref{fig:main}(a) shows that this is true for the most part, but that something unusual happens for $|\Omega| = |\lambda| \gg |\omega|$, where the error remains small. To show this effect in more detail, in Fig. \ref{fig:main}(b) we plot a cross-section of Fig. \ref{fig:main}(a) along the $\Omega = \lambda$ diagonal line. Along this cross-section we see that the estimation error does not simply increase monotonically as the interaction strength $\lambda$ increases. Rather, there is a range of values for which increasing the interaction strength \emph{improves} the estimation error, before the error eventually plateaus for a sufficiently large interaction strength $\lambda$. Fig. \ref{fig:main}(c) shows that this is associated with an optimal sensing time $t_*$ that is also enhanced by increasing $\lambda$. In contrast, a cross-section along the line $\Omega = 2\lambda$ shows that increasing interactions result in a degraded sensitivity, with the error increasing as $\delta\omega_* \sim |\lambda|^{1/2}$ and the optimal sensing time decreasing as $t_* \sim |\lambda|^{-1}$. This is consistent with the usual expectation that increased interactions between spins leads to faster decoherence.

In the next section we will see that, as with the spin-1 example, the reason for the improved sensor performance is the emergence of quantum many-body scars in the parameter regime $|\Omega| = |\lambda| \gg |\omega|$. Before that however, we consider a perturbation to the mixed-field Ising Hamiltonian that was shown by Choi \emph{et al.} to enhance the many-body scars in that parameter regime \cite{Cho-19}. The perturbation is $\hat{H}=\hat{H}_\text{MFI} + \delta\hat{H}$ where: \begin{equation} \delta\hat{H} \equiv \frac{\eta}{4} \sum_{n=0}^{N-1} \sum_{d=2}^{N/2} c_d (\hat{\sigma}_n^x \hat{\sigma}_{n+d}^z + \hat{\sigma}_n^z \hat{\sigma}_{n+d}^x) , \label{eq:delta_H} \end{equation} with $c_d = (\phi^{d-1} - \phi^{-d+1})^{-2}$ and $\phi = (1 + \sqrt{5})/2$ the golden ratio. Choosing $\eta = -0.1\omega$ we see in Fig. \ref{fig:main}(b) that, with this perturbation, a scaling $\delta\omega_* \sim |\lambda|^{-1/2}$ is maintained for a large range of interaction strengths $\lambda$, giving very low error for large interaction strength. Similarly, for $\lambda \gg \omega$ the sensing time scales as $t_* \sim |\lambda|$, i.e., longer sensing times are achieved with stronger interactions. We now explain that the emergence of quantum many-body scars are responsible for this unusual enhancement in sensitivity with increasing interaction strength.
 

\subsection{Quantum many-body scars in the MFI model}\label{sec:MFI_scars}

If $\Omega = \lambda$ the mixed-field Ising Hamiltonian can be rewritten as: \begin{equation} \hat{H}_\text{MFI} = \frac{\omega}{2}\sum_{n=0}^{N-1}\hat{\sigma}_n^x + \lambda \sum_{n=0}^{N-1} \ket{\uparrow_n\uparrow_{n+1}}\bra{\uparrow_n\uparrow_{n+1}} , \end{equation} up to an added constant that just shifts all energies by an equal amount. We can see that for $\Omega = \lambda \gg |\omega|$ states $\ket{\hdots\uparrow\uparrow\hdots}$ with two consecutive $\uparrow$-states have a large energy penalty (alternatively, if $\Omega = -\lambda$ states $\ket{\hdots\downarrow\downarrow\hdots}$ with two consecutive $\downarrow$-states have a large energy penalty). Neglecting these states and performing a rotating wave approximation gives the effective ``PXP Hamiltonian'' \cite{Les-11}: \begin{equation} \hat{H}_\text{PXP} = \hat{P}\left( \frac{\omega}{2}\sum_{n=0}^{N-1}\hat{\sigma}_n^x \right)\hat{P} , \end{equation} where $\hat{P} = \prod_{n=0}^{N-1} ( \mathbb{I} - \ket{\uparrow_n\uparrow_{n+1}}\bra{\uparrow_n\uparrow_{n+1}})$ is a projector that forbids any transitions into states with neighbouring spins in the $\uparrow$-state. The PXP Hamiltonian is known to have a set of quantum many-body scars \cite{Tur-18a}. The scars have approximately equal energy gaps, and a large overlap with the initial N\'{e}el state $\ket{\mathbb{Z}_2}$. This results in long-lived revivals of the initial state, and is the origin of the enhanced sensor performance in the parameter regime $\Omega = \lambda \gg \omega$. However, the revivals are not perfect, and they do eventually decay, corresponding to a finite $t_*$ in our numerical simulations of the sensing experiment. It was shown by Choi \emph{et. al} \cite{Cho-19} that the energy gaps between the scar states can be made almost exactly harmonic, and the revivals almost perfect, by adding the perturbation $\hat{P}(\delta\hat{H})\hat{P}$ to the PXP Hamiltonian, with $\delta\hat{H}$ as given in Eq. \ref{eq:delta_H}. If this perturbation leads to perfect revivals of the initial state we can expect the optimal sensing time $t_*$ to diverge in the PXP-limit. This is the origin of the $t_* \sim |\lambda|$ scaling for $\lambda \gg \omega$ in Fig. \ref{fig:main}(c).

Finally, we note that for the non-interacting model $\lambda=\Omega=\eta=0$, we get the same error $\delta\omega = 1/\sqrt{NtT}$, whether we prepare the spins in the initial N\'{e}el state $\ket{\psi(0)}=\ket{\mathbb{Z}_2}$, or if we choose the fully polarised initial state $\ket{\psi(0)} = \ket{\Downarrow} \equiv \ket{\downarrow\downarrow\downarrow\hdots}$. However, in the parameter regime $\lambda=\Omega\gg\omega$ the N\'{e}el state lives in the scarred subspace while the polarized state does not. The polarized state will therefore thermalise and cannot give improving sensor performance with increasing interaction strength. This is shown in the yellow line in Fig. \ref{fig:main}(b).


\section{Robustness and periodic controls}\label{sec:noise}

The two examples presented in section \ref{sec:spin_1} and section \ref{sec:MFIM} show that quantum sensing is very robust to certain strong interactions, as a result of quantum many-body scarring. However, it is natural to ask how stable such sensing is against other perturbations that might appear in any practical realisation of the scheme. This depends partly on the stability of the scars themselves. Despite some work on this topic \cite{Tur-18b, Khe-19, Lin-20, Sur-20, Shi-20}, there is still much unknown. 



From the sensing perspective, the situation can be compared to attempts to design a quantum sensor by suppressing all interactions between spins $\hat{H}_\text{int} \approx 0$ (see the discussion in section \ref{sec:general}). In that case, even with very good suppression, in realistic experiments there will always be unwanted small perturbations that degrade the sensor performance. Similarly, although our sensing schemes in section \ref{sec:spin_1} and section \ref{sec:MFIM} are perfectly robust to certain strong interactions, in reality there will always be some degradation compared to the ideal case. For $\hat{H}_\text{int} \approx 0$ it is well known that the application of periodic controls can suppress unwanted interactions \cite{Vio-98}, and that this is compatible with the estimation of alternating signals \cite{Deg-17, Tay-08}. Below, we show that this approach can, in principle, be tailored to quantum sensing in a strongly interacting system with many-body scars.





To illustrate the idea we return to our spin-1 Hamiltonian, this time assuming a one-dimensional nearest neighbour interaction $\hat{H}_\text{int} = \lambda \sum_{n}(e^{i\phi}\hat{S}_n^+\hat{S}_{n+1}^- + \text{h.c.})$. Recall that this can be rewritten as $\hat{H}_\text{int} = \lambda \sum_n [ \cos\phi (\hat{S}_n^x\hat{S}_{n+1}^x + \hat{S}_n^y \hat{S}_{n+1}^y) + \sin\phi (\hat{S}_n^x\hat{S}_{n+1}^y - \hat{S}_n^y \hat{S}_{n+1}^x)]$, separating it into its XX-interaction (the $\cos\phi$ term) and DMI contribution (the $\sin\phi$ term). In Eq. \ref{eq:error_fit} we showed (for a slightly different choice of couplings $\lambda_{n,n'}$) that the optimised error $\delta\omega_* \sim \sqrt{|\lambda\cos\phi|}$ is degraded by the strength of the XX-interaction. This seems to indicate that the sensing is not very robust to such a perturbation.

Now consider the unitary operator $\hat{V}_{\pi} \equiv e^{i\pi\hat{S}_0^x}e^{i\pi\hat{S}_1^y}e^{i\pi\hat{S}_2^x}e^{i\pi\hat{S}_3^y}\hdots$, which implements a $\pi$-rotation of each spin around its $x$-axis (for even spin index $n$) or around its $y$-axis (for odd spin index $n$). The $\pi$-pulse $\hat{V}_{\pi}$ has the property that it commutes with the DMI component of $\hat{H}_\text{int}$, but anticommutes with the damaging XX-component, so that $\hat{V}_{\pi}\hat{H}_\text{int}\hat{V}_{\pi} = \lambda \sum_n [ - \cos\phi (\hat{S}_n^x\hat{S}_{n+1}^x + \hat{S}_n^y \hat{S}_{n+1}^y) + \sin\phi (\hat{S}_n^x\hat{S}_{n+1}^y - \hat{S}_n^y \hat{S}_{n+1}^x)]$. We might therefore expect that periodic application of $\hat{V}_{\pi}$ to the spins will have the effect of suppressing the unwanted XX-interaction, while leaving the DMI, to which the sensing is already robust, relatively unaffected. Note, however, that for $\hat{H}_0 = \frac{\omega}{2}\sum_n\hat{S}_n^z$ we have $\hat{V}_{\pi} \hat{H}_0 \hat{V}_{\pi} = -\hat{H}_0$, so that the signal we wish to measure is also suppressed by the pulse. This is a well known issue in quantum sensing with periodic controls, and can be overcome if we modify our scheme to measure a signal that is alternating in time at the pulse frequency, $\hat{H}_0 \to \hat{H}_0(t) = \frac{\omega}{2}\sin(\pi t/\tau) \sum_n \hat{S}_n^z$, where $\tau$ is the time interval between the periodic $\pi$-pulses. With this modification the reversal of sign of $\hat{H}_0$ on the application of a $\pi$-pulse is compensated by the change of sign in the sinusoid, giving an overall accumulation of signal \cite{Tay-08}.


\begin{figure}
  \includegraphics[width=\columnwidth]{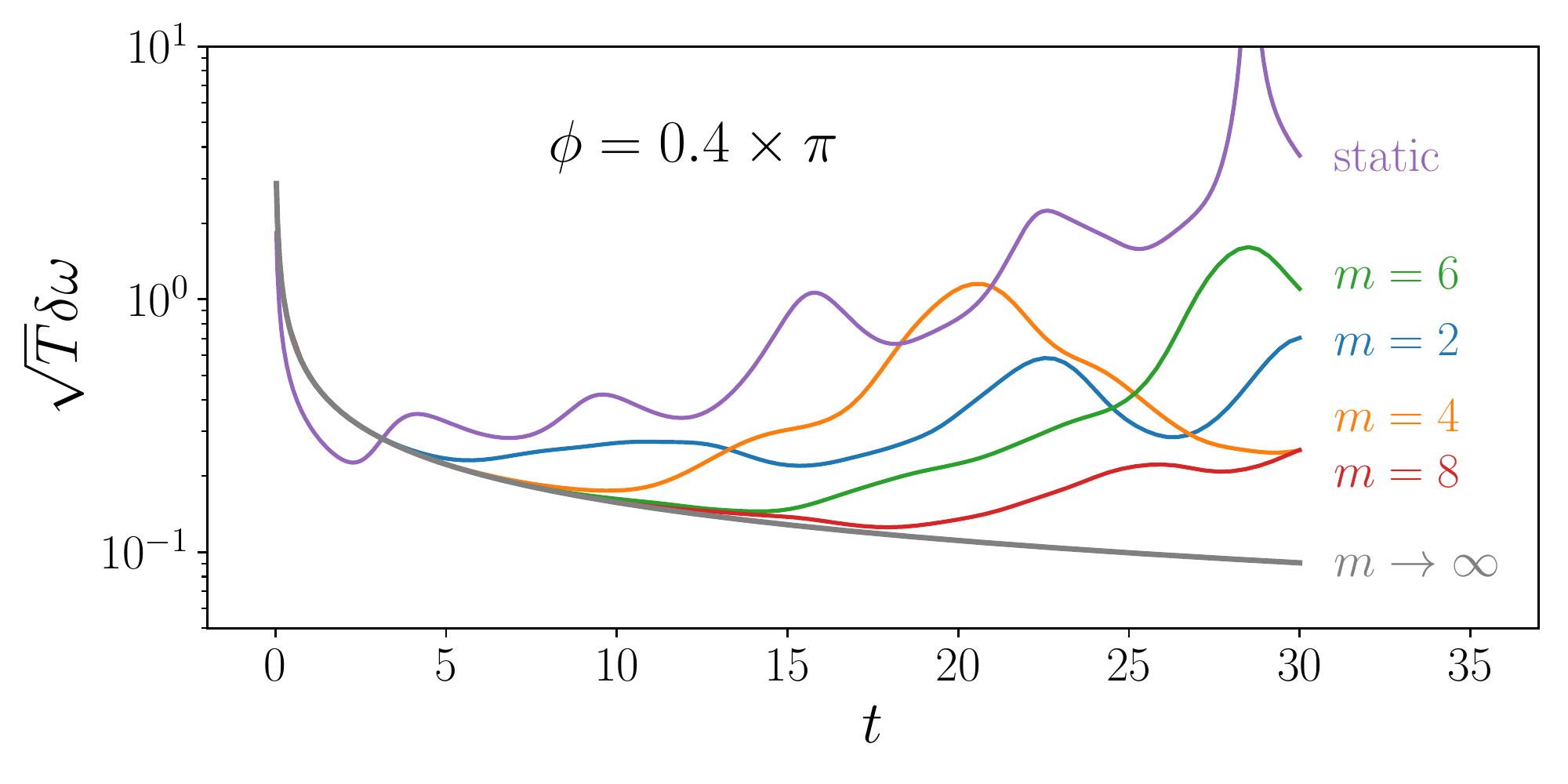}
  \caption{Deviations from $\phi=\pi/2$ (ideal DMI) can degrade the sensitivity in estimating a static parameter $\omega$ (purple line). Periodic $\pi$-pulses can suppress the damaging XX-interaction, extend the optimal sensing time and enhance the sensitivity (in a modified scheme to estimate the amplitude $\omega$ of an alternating signal). Here we plot the error at time $t$ assuming that there have been $m$ periodic $\pi$-pulses $\hat{V}_\pi$ (at the times $t/m, 2t/m,\hdots, t$). [Other parameters: $N=10$, $\lambda = 0.8$, $\phi = 0.4\pi$, $\omega = 1$.]}
  \label{fig:error_with_controls}
\end{figure} 

In Fig. \ref{fig:error_with_controls} we plot the error in estimating the amplitude $\omega$ of the alternating signal at the sensing time $t$, assuming that the $\pi$-pulse $\hat{V}_\pi$ has been applied $m$ times at periodic intervals $\tau = t/m$ during the preceeding dynamics. We see that the periodic controls suppress the damaging XX-interaction, extend the optimal sensing time and enhance the optimal sensitivity. 


Another likely source of noise is inhomogeneity in the local magnetic fields, represented by an added Hamiltonian term $\hat{H}_\Delta = \sum_n \Delta_n \hat{S}_n^z$. Since $\hat{V}_{\pi}\hat{H}_\Delta \hat{V}_{\pi} = -\hat{H}_\Delta$ the $\pi$-pulse is also effective at suppressing such perturbations. In the limit of high-frequency periodic control, $m\to\infty$, the optimal sensing time diverges and the estimation error is $\delta\omega = 2 / \sqrt{\pi^2 NtT}$, as shown in Appendix \ref{app:noise}.



We note that the simple periodic control considered here does not suppress \emph{next}-nearest neighbour XX-interactions or interactions of the form $\hat{H}_Z = \sum_{n,n'}g_{n,n'}\hat{S}_n^z\hat{S}_{n'}^z$, if they are included. Also, pulse errors have not been taken into account in the analysis above, although they may be significant if many pulses $m \gg 1$ are applied. One possible approach to these issues might be to develop a more sophisticated pulse sequence that is robust to more generic noise and to pulse errors \cite{DeL-10}. Another promising approach -- recently demonstated experimentally -- is to employ periodic driving to stabilise the scars and the associated long-lived oscillations by the creation of a robust discrete time-crystal-like phase \cite{Blu-21, Mas-21}. Further work is required to determine if such a robust non-equilibrium phase can be exploited for quantum sensing.

\section{Conclusion}


Quantum many-body scars are special eigenstates of a non-integrable many-body system that, for certain initial states, can prevent or slow down thermalisation. Since this is associated with long coherence times, scars can be exploited for quantum sensing. In this paper we have demonstrated this for two example models: a spin-1 DMI model, and a spin-1/2 MFI model.



Although the two examples appear to be very different, there are some interesting similarities in the structure of their scar subspaces \cite{Sch-19}. Recall that the scar states in the spin-1 DMI model are the Dicke states $\{\ket{\Psi(s)}\}_{s=0}^N$ (defined in Eq. \ref{eq:dicke_states}) with harmonic energy gaps, resulting in SU(2)-spin dynamics in the Dicke subspace. For the spin-1/2 PXP-model, it was shown in Ref. \cite{Tur-18a} that the dynamics in the scar subspace can be though of as approximate SU(2)-spin dynamics, with the scar states playing the role of the Dicke subspace. The perturbation introduced at the end of section \ref{sec:MFIM} improves the approximation, so that the dynamics are almost exactly like those of an SU(2) spin. Whether this SU(2) structure is an essential feature of quantum sensing via many-body scars, or if examples exist without this feature (but still satisfying our criteria in section \ref{sec:general}) is, to the best of our knowledge, an open question.

In both examples we assumed an initial product state of the $N$ probe particles. However, it is well known that for non-interacting systems, entangled initial states such as spin squeezed states can give an enhanced sensitivity compared to separable initial states \cite{Gio-04}. This is also possible for strongly interacting systems with many-body scars, as we show in Appendix \ref{app:squeezing} for the spin-1 DMI example.

Robustness to perturbations is undoubtedly an important topic of further research, if quantum sensors exploiting scars are to become a useful technology. As discussed in section \ref{sec:noise}, periodic controls may offer a route to such robustness.

\begin{acknowledgments}
  The author thanks G. Kells for discussions and for helpful comments on the manuscript. This work was funded by Science Foundation Ireland through Career Development Award 15/CDA/3240. The author also wishes to acknowledge the DJEI/DES/SFI/HEA Irish Centre for High-End Computing (ICHEC) for the provision of computational facilities.
\end{acknowledgments}


\bibliography{/Users/dooleysh/Google_Drive/physics/papers/bibtex_library/refs}

\begin{thebibliography}{69}%
\makeatletter
\providecommand \@ifxundefined [1]{%
 \@ifx{#1\undefined}
}%
\providecommand \@ifnum [1]{%
 \ifnum #1\expandafter \@firstoftwo
 \else \expandafter \@secondoftwo
 \fi
}%
\providecommand \@ifx [1]{%
 \ifx #1\expandafter \@firstoftwo
 \else \expandafter \@secondoftwo
 \fi
}%
\providecommand \natexlab [1]{#1}%
\providecommand \enquote  [1]{``#1''}%
\providecommand \bibnamefont  [1]{#1}%
\providecommand \bibfnamefont [1]{#1}%
\providecommand \citenamefont [1]{#1}%
\providecommand \href@noop [0]{\@secondoftwo}%
\providecommand \href [0]{\begingroup \@sanitize@url \@href}%
\providecommand \@href[1]{\@@startlink{#1}\@@href}%
\providecommand \@@href[1]{\endgroup#1\@@endlink}%
\providecommand \@sanitize@url [0]{\catcode `\\12\catcode `\$12\catcode
  `\&12\catcode `\#12\catcode `\^12\catcode `\_12\catcode `\%12\relax}%
\providecommand \@@startlink[1]{}%
\providecommand \@@endlink[0]{}%
\providecommand \url  [0]{\begingroup\@sanitize@url \@url }%
\providecommand \@url [1]{\endgroup\@href {#1}{\urlprefix }}%
\providecommand \urlprefix  [0]{URL }%
\providecommand \Eprint [0]{\href }%
\providecommand \doibase [0]{http://dx.doi.org/}%
\providecommand \selectlanguage [0]{\@gobble}%
\providecommand \bibinfo  [0]{\@secondoftwo}%
\providecommand \bibfield  [0]{\@secondoftwo}%
\providecommand \translation [1]{[#1]}%
\providecommand \BibitemOpen [0]{}%
\providecommand \bibitemStop [0]{}%
\providecommand \bibitemNoStop [0]{.\EOS\space}%
\providecommand \EOS [0]{\spacefactor3000\relax}%
\providecommand \BibitemShut  [1]{\csname bibitem#1\endcsname}%
\let\auto@bib@innerbib\@empty
\bibitem [{\citenamefont {Giovannetti}\ \emph {et~al.}(2011)\citenamefont
  {Giovannetti}, \citenamefont {Lloyd},\ and\ \citenamefont
  {Maccone}}]{Gio-11}%
  \BibitemOpen
  \bibfield  {author} {\bibinfo {author} {\bibfnamefont {V.}~\bibnamefont
  {Giovannetti}}, \bibinfo {author} {\bibfnamefont {S.}~\bibnamefont {Lloyd}},
  \ and\ \bibinfo {author} {\bibfnamefont {L.}~\bibnamefont {Maccone}},\
  }\bibfield  {title} {\enquote {\bibinfo {title} {Advances in quantum
  metrology},}\ }\href@noop {} {\bibfield  {journal} {\bibinfo  {journal}
  {Nature Photonics}\ }\textbf {\bibinfo {volume} {5}},\ \bibinfo {pages}
  {222--229} (\bibinfo {year} {2011})}\BibitemShut {NoStop}%
\bibitem [{\citenamefont {Degen}\ \emph {et~al.}(2017)\citenamefont {Degen},
  \citenamefont {Reinhard},\ and\ \citenamefont {Cappellaro}}]{Deg-17}%
  \BibitemOpen
  \bibfield  {author} {\bibinfo {author} {\bibfnamefont {C.~L.}\ \bibnamefont
  {Degen}}, \bibinfo {author} {\bibfnamefont {F.}~\bibnamefont {Reinhard}}, \
  and\ \bibinfo {author} {\bibfnamefont {P.}~\bibnamefont {Cappellaro}},\
  }\bibfield  {title} {\enquote {\bibinfo {title} {Quantum sensing},}\ }\href
  {\doibase 10.1103/RevModPhys.89.035002} {\bibfield  {journal} {\bibinfo
  {journal} {Rev. Mod. Phys.}\ }\textbf {\bibinfo {volume} {89}},\ \bibinfo
  {pages} {035002} (\bibinfo {year} {2017})}\BibitemShut {NoStop}%
\bibitem [{\citenamefont {Pezz\`e}\ \emph {et~al.}(2018)\citenamefont
  {Pezz\`e}, \citenamefont {Smerzi}, \citenamefont {Oberthaler}, \citenamefont
  {Schmied},\ and\ \citenamefont {Treutlein}}]{Pez-18}%
  \BibitemOpen
  \bibfield  {author} {\bibinfo {author} {\bibfnamefont {Luca}\ \bibnamefont
  {Pezz\`e}}, \bibinfo {author} {\bibfnamefont {Augusto}\ \bibnamefont
  {Smerzi}}, \bibinfo {author} {\bibfnamefont {Markus~K.}\ \bibnamefont
  {Oberthaler}}, \bibinfo {author} {\bibfnamefont {Roman}\ \bibnamefont
  {Schmied}}, \ and\ \bibinfo {author} {\bibfnamefont {Philipp}\ \bibnamefont
  {Treutlein}},\ }\bibfield  {title} {\enquote {\bibinfo {title} {Quantum
  metrology with nonclassical states of atomic ensembles},}\ }\href {\doibase
  10.1103/RevModPhys.90.035005} {\bibfield  {journal} {\bibinfo  {journal}
  {Rev. Mod. Phys.}\ }\textbf {\bibinfo {volume} {90}},\ \bibinfo {pages}
  {035005} (\bibinfo {year} {2018})}\BibitemShut {NoStop}%
\bibitem [{\citenamefont {Budker}\ and\ \citenamefont
  {Romalis}(2007)}]{Bud-07}%
  \BibitemOpen
  \bibfield  {author} {\bibinfo {author} {\bibfnamefont {Dmitry}\ \bibnamefont
  {Budker}}\ and\ \bibinfo {author} {\bibfnamefont {Michael}\ \bibnamefont
  {Romalis}},\ }\bibfield  {title} {\enquote {\bibinfo {title} {Optical
  magnetometry},}\ }\href@noop {} {\bibfield  {journal} {\bibinfo  {journal}
  {Nature Physics}\ }\textbf {\bibinfo {volume} {3}} (\bibinfo {year}
  {2007})}\BibitemShut {NoStop}%
\bibitem [{\citenamefont {Taylor}\ \emph {et~al.}(2008)\citenamefont {Taylor},
  \citenamefont {Cappellaro}, \citenamefont {Childress}, \citenamefont {Jiang},
  \citenamefont {Budker}, \citenamefont {Hemmer}, \citenamefont {Yacoby},
  \citenamefont {Walsworth},\ and\ \citenamefont {Lukin}}]{Tay-08}%
  \BibitemOpen
  \bibfield  {author} {\bibinfo {author} {\bibfnamefont {JM}~\bibnamefont
  {Taylor}}, \bibinfo {author} {\bibfnamefont {P}~\bibnamefont {Cappellaro}},
  \bibinfo {author} {\bibfnamefont {L}~\bibnamefont {Childress}}, \bibinfo
  {author} {\bibfnamefont {L}~\bibnamefont {Jiang}}, \bibinfo {author}
  {\bibfnamefont {D}~\bibnamefont {Budker}}, \bibinfo {author} {\bibfnamefont
  {PR}~\bibnamefont {Hemmer}}, \bibinfo {author} {\bibfnamefont
  {A}~\bibnamefont {Yacoby}}, \bibinfo {author} {\bibfnamefont {R}~\bibnamefont
  {Walsworth}}, \ and\ \bibinfo {author} {\bibfnamefont {MD}~\bibnamefont
  {Lukin}},\ }\bibfield  {title} {\enquote {\bibinfo {title} {High-sensitivity
  diamond magnetometer with nanoscale resolution},}\ }\href@noop {} {\bibfield
  {journal} {\bibinfo  {journal} {Nature Physics}\ }\textbf {\bibinfo {volume}
  {4}},\ \bibinfo {pages} {810--816} (\bibinfo {year} {2008})}\BibitemShut
  {NoStop}%
\bibitem [{\citenamefont {Tanaka}\ \emph {et~al.}(2015)\citenamefont {Tanaka},
  \citenamefont {Knott}, \citenamefont {Matsuzaki}, \citenamefont {Dooley},
  \citenamefont {Yamaguchi}, \citenamefont {Munro},\ and\ \citenamefont
  {Saito}}]{Tan-15}%
  \BibitemOpen
  \bibfield  {author} {\bibinfo {author} {\bibfnamefont {Tohru}\ \bibnamefont
  {Tanaka}}, \bibinfo {author} {\bibfnamefont {Paul}\ \bibnamefont {Knott}},
  \bibinfo {author} {\bibfnamefont {Yuichiro}\ \bibnamefont {Matsuzaki}},
  \bibinfo {author} {\bibfnamefont {Shane}\ \bibnamefont {Dooley}}, \bibinfo
  {author} {\bibfnamefont {Hiroshi}\ \bibnamefont {Yamaguchi}}, \bibinfo
  {author} {\bibfnamefont {William~J.}\ \bibnamefont {Munro}}, \ and\ \bibinfo
  {author} {\bibfnamefont {Shiro}\ \bibnamefont {Saito}},\ }\bibfield  {title}
  {\enquote {\bibinfo {title} {Proposed robust entanglement-based magnetic
  field sensor beyond the standard quantum limit},}\ }\href {\doibase
  10.1103/PhysRevLett.115.170801} {\bibfield  {journal} {\bibinfo  {journal}
  {Phys. Rev. Lett.}\ }\textbf {\bibinfo {volume} {115}},\ \bibinfo {pages}
  {170801} (\bibinfo {year} {2015})}\BibitemShut {NoStop}%
\bibitem [{\citenamefont {Dolde}\ \emph {et~al.}(2011)\citenamefont {Dolde},
  \citenamefont {Fedder}, \citenamefont {Doherty}, \citenamefont {N{\"o}bauer},
  \citenamefont {Rempp}, \citenamefont {Balasubramanian}, \citenamefont {Wolf},
  \citenamefont {Reinhard}, \citenamefont {Hollenberg}, \citenamefont {Jelezko}
  \emph {et~al.}}]{Dol-11}%
  \BibitemOpen
  \bibfield  {author} {\bibinfo {author} {\bibfnamefont {F}~\bibnamefont
  {Dolde}}, \bibinfo {author} {\bibfnamefont {H}~\bibnamefont {Fedder}},
  \bibinfo {author} {\bibfnamefont {MW}~\bibnamefont {Doherty}}, \bibinfo
  {author} {\bibfnamefont {T}~\bibnamefont {N{\"o}bauer}}, \bibinfo {author}
  {\bibfnamefont {F}~\bibnamefont {Rempp}}, \bibinfo {author} {\bibfnamefont
  {G}~\bibnamefont {Balasubramanian}}, \bibinfo {author} {\bibfnamefont
  {T}~\bibnamefont {Wolf}}, \bibinfo {author} {\bibfnamefont {F}~\bibnamefont
  {Reinhard}}, \bibinfo {author} {\bibfnamefont {LCL}\ \bibnamefont
  {Hollenberg}}, \bibinfo {author} {\bibfnamefont {F}~\bibnamefont {Jelezko}},
  \emph {et~al.},\ }\bibfield  {title} {\enquote {\bibinfo {title}
  {Electric-field sensing using single diamond spins},}\ }\href@noop {}
  {\bibfield  {journal} {\bibinfo  {journal} {Nature Physics}\ }\textbf
  {\bibinfo {volume} {7}},\ \bibinfo {pages} {459} (\bibinfo {year}
  {2011})}\BibitemShut {NoStop}%
\bibitem [{\citenamefont {Facon}\ \emph {et~al.}(2016)\citenamefont {Facon},
  \citenamefont {Dietsche}, \citenamefont {Grosso}, \citenamefont {Haroche},
  \citenamefont {Raimond}, \citenamefont {Brune},\ and\ \citenamefont
  {Gleyzes}}]{Fac-16}%
  \BibitemOpen
  \bibfield  {author} {\bibinfo {author} {\bibfnamefont {Adrien}\ \bibnamefont
  {Facon}}, \bibinfo {author} {\bibfnamefont {Eva-Katharina}\ \bibnamefont
  {Dietsche}}, \bibinfo {author} {\bibfnamefont {Dorian}\ \bibnamefont
  {Grosso}}, \bibinfo {author} {\bibfnamefont {Serge}\ \bibnamefont {Haroche}},
  \bibinfo {author} {\bibfnamefont {Jean-Michel}\ \bibnamefont {Raimond}},
  \bibinfo {author} {\bibfnamefont {Michel}\ \bibnamefont {Brune}}, \ and\
  \bibinfo {author} {\bibfnamefont {S{\'e}bastien}\ \bibnamefont {Gleyzes}},\
  }\bibfield  {title} {\enquote {\bibinfo {title} {A sensitive electrometer
  based on a {R}ydberg atom in a {S}chr{\"o}dinger-cat state},}\ }\href@noop {}
  {\bibfield  {journal} {\bibinfo  {journal} {Nature}\ }\textbf {\bibinfo
  {volume} {535}},\ \bibinfo {pages} {262--265} (\bibinfo {year}
  {2016})}\BibitemShut {NoStop}%
\bibitem [{\citenamefont {Ludlow}\ \emph {et~al.}(2015)\citenamefont {Ludlow},
  \citenamefont {Boyd}, \citenamefont {Ye}, \citenamefont {Peik},\ and\
  \citenamefont {Schmidt}}]{Lud-15}%
  \BibitemOpen
  \bibfield  {author} {\bibinfo {author} {\bibfnamefont {Andrew~D.}\
  \bibnamefont {Ludlow}}, \bibinfo {author} {\bibfnamefont {Martin~M.}\
  \bibnamefont {Boyd}}, \bibinfo {author} {\bibfnamefont {Jun}\ \bibnamefont
  {Ye}}, \bibinfo {author} {\bibfnamefont {E.}~\bibnamefont {Peik}}, \ and\
  \bibinfo {author} {\bibfnamefont {P.~O.}\ \bibnamefont {Schmidt}},\
  }\bibfield  {title} {\enquote {\bibinfo {title} {Optical atomic clocks},}\
  }\href {\doibase 10.1103/RevModPhys.87.637} {\bibfield  {journal} {\bibinfo
  {journal} {Rev. Mod. Phys.}\ }\textbf {\bibinfo {volume} {87}},\ \bibinfo
  {pages} {637--701} (\bibinfo {year} {2015})}\BibitemShut {NoStop}%
\bibitem [{\citenamefont {Caves}(1981)}]{Cav-81}%
  \BibitemOpen
  \bibfield  {author} {\bibinfo {author} {\bibfnamefont {Carlton~M}\
  \bibnamefont {Caves}},\ }\bibfield  {title} {\enquote {\bibinfo {title}
  {Quantum-mechanical noise in an interferometer},}\ }\href@noop {} {\bibfield
  {journal} {\bibinfo  {journal} {Physical Review D}\ }\textbf {\bibinfo
  {volume} {23}},\ \bibinfo {pages} {1693} (\bibinfo {year}
  {1981})}\BibitemShut {NoStop}%
\bibitem [{\citenamefont {Aasi}\ \emph {et~al.}(2013)\citenamefont {Aasi},
  \citenamefont {Abadie}, \citenamefont {Abbott}, \citenamefont {Abbott},
  \citenamefont {Abernathy}, \citenamefont {Adhikari}, \citenamefont {Ajith},
  \citenamefont {Anderson}, \citenamefont {Arai}, \citenamefont {Araya} \emph
  {et~al.}}]{Aas-13}%
  \BibitemOpen
  \bibfield  {author} {\bibinfo {author} {\bibfnamefont {J}~\bibnamefont
  {Aasi}}, \bibinfo {author} {\bibfnamefont {J}~\bibnamefont {Abadie}},
  \bibinfo {author} {\bibfnamefont {BP}~\bibnamefont {Abbott}}, \bibinfo
  {author} {\bibfnamefont {R}~\bibnamefont {Abbott}}, \bibinfo {author}
  {\bibfnamefont {MR}~\bibnamefont {Abernathy}}, \bibinfo {author}
  {\bibfnamefont {RX}~\bibnamefont {Adhikari}}, \bibinfo {author}
  {\bibfnamefont {P}~\bibnamefont {Ajith}}, \bibinfo {author} {\bibfnamefont
  {SB}~\bibnamefont {Anderson}}, \bibinfo {author} {\bibfnamefont
  {K}~\bibnamefont {Arai}}, \bibinfo {author} {\bibfnamefont {MC}~\bibnamefont
  {Araya}},  \emph {et~al.},\ }\bibfield  {title} {\enquote {\bibinfo {title}
  {Enhanced sensitivity of the {LIGO} gravitational wave detector by using
  squeezed states of light},}\ }\href@noop {} {\bibfield  {journal} {\bibinfo
  {journal} {Nature Photonics}\ }\textbf {\bibinfo {volume} {7}},\ \bibinfo
  {pages} {613--619} (\bibinfo {year} {2013})}\BibitemShut {NoStop}%
\bibitem [{\citenamefont {Yurke}\ \emph {et~al.}(1986)\citenamefont {Yurke},
  \citenamefont {McCall},\ and\ \citenamefont {Klauder}}]{Yur-86b}%
  \BibitemOpen
  \bibfield  {author} {\bibinfo {author} {\bibfnamefont {Bernard}\ \bibnamefont
  {Yurke}}, \bibinfo {author} {\bibfnamefont {Samuel~L}\ \bibnamefont
  {McCall}}, \ and\ \bibinfo {author} {\bibfnamefont {John~R}\ \bibnamefont
  {Klauder}},\ }\bibfield  {title} {\enquote {\bibinfo {title} {{SU}(2) and
  {SU}(1, 1) interferometers},}\ }\href@noop {} {\bibfield  {journal} {\bibinfo
   {journal} {Physical Review A}\ }\textbf {\bibinfo {volume} {33}},\ \bibinfo
  {pages} {4033} (\bibinfo {year} {1986})}\BibitemShut {NoStop}%
\bibitem [{\citenamefont {Lee}\ \emph {et~al.}(2002)\citenamefont {Lee},
  \citenamefont {Kok},\ and\ \citenamefont {Dowling}}]{Lee-02}%
  \BibitemOpen
  \bibfield  {author} {\bibinfo {author} {\bibfnamefont {Hwang}\ \bibnamefont
  {Lee}}, \bibinfo {author} {\bibfnamefont {Pieter}\ \bibnamefont {Kok}}, \
  and\ \bibinfo {author} {\bibfnamefont {Jonathan~P.}\ \bibnamefont
  {Dowling}},\ }\bibfield  {title} {\enquote {\bibinfo {title} {A quantum
  {R}osetta stone for interferometry},}\ }\href {\doibase
  10.1080/0950034021000011536} {\bibfield  {journal} {\bibinfo  {journal}
  {Journal of Modern Optics}\ }\textbf {\bibinfo {volume} {49}},\ \bibinfo
  {pages} {2325--2338} (\bibinfo {year} {2002})}\BibitemShut {NoStop}%
\bibitem [{\citenamefont {Dooley}\ \emph {et~al.}(2018)\citenamefont {Dooley},
  \citenamefont {Hanks}, \citenamefont {Nakayama}, \citenamefont {Munro},\ and\
  \citenamefont {Nemoto}}]{Doo-18a}%
  \BibitemOpen
  \bibfield  {author} {\bibinfo {author} {\bibfnamefont {Shane}\ \bibnamefont
  {Dooley}}, \bibinfo {author} {\bibfnamefont {Michael}\ \bibnamefont {Hanks}},
  \bibinfo {author} {\bibfnamefont {Shojun}\ \bibnamefont {Nakayama}}, \bibinfo
  {author} {\bibfnamefont {William~J}\ \bibnamefont {Munro}}, \ and\ \bibinfo
  {author} {\bibfnamefont {Kae}\ \bibnamefont {Nemoto}},\ }\bibfield  {title}
  {\enquote {\bibinfo {title} {Robust quantum sensing with strongly interacting
  probe systems},}\ }\href@noop {} {\bibfield  {journal} {\bibinfo  {journal}
  {npj Quantum Information}\ }\textbf {\bibinfo {volume} {4}},\ \bibinfo
  {pages} {1--7} (\bibinfo {year} {2018})}\BibitemShut {NoStop}%
\bibitem [{\citenamefont {Choi}\ \emph {et~al.}(2017)\citenamefont {Choi},
  \citenamefont {Yao},\ and\ \citenamefont {Lukin}}]{Cho-17}%
  \BibitemOpen
  \bibfield  {author} {\bibinfo {author} {\bibfnamefont {Soonwon}\ \bibnamefont
  {Choi}}, \bibinfo {author} {\bibfnamefont {Norman~Y}\ \bibnamefont {Yao}}, \
  and\ \bibinfo {author} {\bibfnamefont {Mikhail~D}\ \bibnamefont {Lukin}},\
  }\bibfield  {title} {\enquote {\bibinfo {title} {Quantum metrology based on
  strongly correlated matter},}\ }\href@noop {} {\bibfield  {journal} {\bibinfo
   {journal} {arXiv preprint arXiv:1801.00042}\ } (\bibinfo {year}
  {2017})}\BibitemShut {NoStop}%
\bibitem [{\citenamefont {Raghunandan}\ \emph {et~al.}(2018)\citenamefont
  {Raghunandan}, \citenamefont {Wrachtrup},\ and\ \citenamefont
  {Weimer}}]{Rag-18}%
  \BibitemOpen
  \bibfield  {author} {\bibinfo {author} {\bibfnamefont {Meghana}\ \bibnamefont
  {Raghunandan}}, \bibinfo {author} {\bibfnamefont {J\"org}\ \bibnamefont
  {Wrachtrup}}, \ and\ \bibinfo {author} {\bibfnamefont {Hendrik}\ \bibnamefont
  {Weimer}},\ }\bibfield  {title} {\enquote {\bibinfo {title} {High-density
  quantum sensing with dissipative first order transitions},}\ }\href {\doibase
  10.1103/PhysRevLett.120.150501} {\bibfield  {journal} {\bibinfo  {journal}
  {Phys. Rev. Lett.}\ }\textbf {\bibinfo {volume} {120}},\ \bibinfo {pages}
  {150501} (\bibinfo {year} {2018})}\BibitemShut {NoStop}%
\bibitem [{\citenamefont {Zhou}\ \emph {et~al.}(2020)\citenamefont {Zhou},
  \citenamefont {Choi}, \citenamefont {Choi}, \citenamefont {Landig},
  \citenamefont {Douglas}, \citenamefont {Isoya}, \citenamefont {Jelezko},
  \citenamefont {Onoda}, \citenamefont {Sumiya}, \citenamefont {Cappellaro},
  \citenamefont {Knowles}, \citenamefont {Park},\ and\ \citenamefont
  {Lukin}}]{Zho-20}%
  \BibitemOpen
  \bibfield  {author} {\bibinfo {author} {\bibfnamefont {Hengyun}\ \bibnamefont
  {Zhou}}, \bibinfo {author} {\bibfnamefont {Joonhee}\ \bibnamefont {Choi}},
  \bibinfo {author} {\bibfnamefont {Soonwon}\ \bibnamefont {Choi}}, \bibinfo
  {author} {\bibfnamefont {Renate}\ \bibnamefont {Landig}}, \bibinfo {author}
  {\bibfnamefont {Alexander~M.}\ \bibnamefont {Douglas}}, \bibinfo {author}
  {\bibfnamefont {Junichi}\ \bibnamefont {Isoya}}, \bibinfo {author}
  {\bibfnamefont {Fedor}\ \bibnamefont {Jelezko}}, \bibinfo {author}
  {\bibfnamefont {Shinobu}\ \bibnamefont {Onoda}}, \bibinfo {author}
  {\bibfnamefont {Hitoshi}\ \bibnamefont {Sumiya}}, \bibinfo {author}
  {\bibfnamefont {Paola}\ \bibnamefont {Cappellaro}}, \bibinfo {author}
  {\bibfnamefont {Helena~S.}\ \bibnamefont {Knowles}}, \bibinfo {author}
  {\bibfnamefont {Hongkun}\ \bibnamefont {Park}}, \ and\ \bibinfo {author}
  {\bibfnamefont {Mikhail~D.}\ \bibnamefont {Lukin}},\ }\bibfield  {title}
  {\enquote {\bibinfo {title} {Quantum metrology with strongly interacting spin
  systems},}\ }\href {\doibase 10.1103/PhysRevX.10.031003} {\bibfield
  {journal} {\bibinfo  {journal} {Phys. Rev. X}\ }\textbf {\bibinfo {volume}
  {10}},\ \bibinfo {pages} {031003} (\bibinfo {year} {2020})}\BibitemShut
  {NoStop}%
\bibitem [{\citenamefont {Bernien}\ \emph {et~al.}(2017)\citenamefont
  {Bernien}, \citenamefont {Schwartz}, \citenamefont {Keesling}, \citenamefont
  {Levine}, \citenamefont {Omran}, \citenamefont {Pichler}, \citenamefont
  {Choi}, \citenamefont {Zibrov}, \citenamefont {Endres}, \citenamefont
  {Greiner}, \citenamefont {Vuleti{\'c}},\ and\ \citenamefont
  {Lukin}}]{Ber-17}%
  \BibitemOpen
  \bibfield  {author} {\bibinfo {author} {\bibfnamefont {Hannes}\ \bibnamefont
  {Bernien}}, \bibinfo {author} {\bibfnamefont {Sylvain}\ \bibnamefont
  {Schwartz}}, \bibinfo {author} {\bibfnamefont {Alexander}\ \bibnamefont
  {Keesling}}, \bibinfo {author} {\bibfnamefont {Harry}\ \bibnamefont
  {Levine}}, \bibinfo {author} {\bibfnamefont {Ahmed}\ \bibnamefont {Omran}},
  \bibinfo {author} {\bibfnamefont {Hannes}\ \bibnamefont {Pichler}}, \bibinfo
  {author} {\bibfnamefont {Soonwon}\ \bibnamefont {Choi}}, \bibinfo {author}
  {\bibfnamefont {Alexander~S.}\ \bibnamefont {Zibrov}}, \bibinfo {author}
  {\bibfnamefont {Manuel}\ \bibnamefont {Endres}}, \bibinfo {author}
  {\bibfnamefont {Markus}\ \bibnamefont {Greiner}}, \bibinfo {author}
  {\bibfnamefont {Vladan}\ \bibnamefont {Vuleti{\'c}}}, \ and\ \bibinfo
  {author} {\bibfnamefont {Mikhail~D.}\ \bibnamefont {Lukin}},\ }\bibfield
  {title} {\enquote {\bibinfo {title} {Probing many-body dynamics on a 51-atom
  quantum simulator},}\ }\href {https://doi.org/10.1038/nature24622} {\bibfield
   {journal} {\bibinfo  {journal} {Nature}\ }\textbf {\bibinfo {volume}
  {551}},\ \bibinfo {pages} {579 EP --} (\bibinfo {year} {2017})}\BibitemShut
  {NoStop}%
\bibitem [{\citenamefont {Turner}\ \emph
  {et~al.}(2018{\natexlab{a}})\citenamefont {Turner}, \citenamefont
  {Michailidis}, \citenamefont {Abanin}, \citenamefont {Serbyn},\ and\
  \citenamefont {Papi{\'c}}}]{Tur-18a}%
  \BibitemOpen
  \bibfield  {author} {\bibinfo {author} {\bibfnamefont {C.~J.}\ \bibnamefont
  {Turner}}, \bibinfo {author} {\bibfnamefont {A.~A.}\ \bibnamefont
  {Michailidis}}, \bibinfo {author} {\bibfnamefont {D.~A.}\ \bibnamefont
  {Abanin}}, \bibinfo {author} {\bibfnamefont {M.}~\bibnamefont {Serbyn}}, \
  and\ \bibinfo {author} {\bibfnamefont {Z.}~\bibnamefont {Papi{\'c}}},\
  }\bibfield  {title} {\enquote {\bibinfo {title} {Weak ergodicity breaking
  from quantum many-body scars},}\ }\href {\doibase 10.1038/s41567-018-0137-5}
  {\bibfield  {journal} {\bibinfo  {journal} {Nature Physics}\ }\textbf
  {\bibinfo {volume} {14}},\ \bibinfo {pages} {745--749} (\bibinfo {year}
  {2018}{\natexlab{a}})}\BibitemShut {NoStop}%
\bibitem [{\citenamefont {Serbyn}\ \emph {et~al.}(2020)\citenamefont {Serbyn},
  \citenamefont {Abanin},\ and\ \citenamefont {Papi{\'c}}}]{Ser-20}%
  \BibitemOpen
  \bibfield  {author} {\bibinfo {author} {\bibfnamefont {Maksym}\ \bibnamefont
  {Serbyn}}, \bibinfo {author} {\bibfnamefont {Dmitry~A.}\ \bibnamefont
  {Abanin}}, \ and\ \bibinfo {author} {\bibfnamefont {Zlatko}\ \bibnamefont
  {Papi{\'c}}},\ }\href@noop {} {\enquote {\bibinfo {title} {Quantum many-body
  scars and weak breaking of ergodicity},}\ } (\bibinfo {year} {2020}),\
  \Eprint {http://arxiv.org/abs/2011.09486} {arXiv:2011.09486 [quant-ph]}
  \BibitemShut {NoStop}%
\bibitem [{\citenamefont {Moudgalya}\ \emph
  {et~al.}(2018{\natexlab{a}})\citenamefont {Moudgalya}, \citenamefont
  {Rachel}, \citenamefont {Bernevig},\ and\ \citenamefont
  {Regnault}}]{Mou-18a}%
  \BibitemOpen
  \bibfield  {author} {\bibinfo {author} {\bibfnamefont {Sanjay}\ \bibnamefont
  {Moudgalya}}, \bibinfo {author} {\bibfnamefont {Stephan}\ \bibnamefont
  {Rachel}}, \bibinfo {author} {\bibfnamefont {B.~Andrei}\ \bibnamefont
  {Bernevig}}, \ and\ \bibinfo {author} {\bibfnamefont {Nicolas}\ \bibnamefont
  {Regnault}},\ }\bibfield  {title} {\enquote {\bibinfo {title} {Exact excited
  states of nonintegrable models},}\ }\href {\doibase
  10.1103/PhysRevB.98.235155} {\bibfield  {journal} {\bibinfo  {journal} {Phys.
  Rev. B}\ }\textbf {\bibinfo {volume} {98}},\ \bibinfo {pages} {235155}
  (\bibinfo {year} {2018}{\natexlab{a}})}\BibitemShut {NoStop}%
\bibitem [{\citenamefont {Moudgalya}\ \emph
  {et~al.}(2018{\natexlab{b}})\citenamefont {Moudgalya}, \citenamefont
  {Regnault},\ and\ \citenamefont {Bernevig}}]{Mou-18b}%
  \BibitemOpen
  \bibfield  {author} {\bibinfo {author} {\bibfnamefont {Sanjay}\ \bibnamefont
  {Moudgalya}}, \bibinfo {author} {\bibfnamefont {Nicolas}\ \bibnamefont
  {Regnault}}, \ and\ \bibinfo {author} {\bibfnamefont {B.~Andrei}\
  \bibnamefont {Bernevig}},\ }\bibfield  {title} {\enquote {\bibinfo {title}
  {Entanglement of exact excited states of affleck-kennedy-lieb-tasaki models:
  Exact results, many-body scars, and violation of the strong eigenstate
  thermalization hypothesis},}\ }\href {\doibase 10.1103/PhysRevB.98.235156}
  {\bibfield  {journal} {\bibinfo  {journal} {Phys. Rev. B}\ }\textbf {\bibinfo
  {volume} {98}},\ \bibinfo {pages} {235156} (\bibinfo {year}
  {2018}{\natexlab{b}})}\BibitemShut {NoStop}%
\bibitem [{\citenamefont {Ok}\ \emph {et~al.}(2019)\citenamefont {Ok},
  \citenamefont {Choo}, \citenamefont {Mudry}, \citenamefont {Castelnovo},
  \citenamefont {Chamon},\ and\ \citenamefont {Neupert}}]{Ok-19}%
  \BibitemOpen
  \bibfield  {author} {\bibinfo {author} {\bibfnamefont {Seulgi}\ \bibnamefont
  {Ok}}, \bibinfo {author} {\bibfnamefont {Kenny}\ \bibnamefont {Choo}},
  \bibinfo {author} {\bibfnamefont {Christopher}\ \bibnamefont {Mudry}},
  \bibinfo {author} {\bibfnamefont {Claudio}\ \bibnamefont {Castelnovo}},
  \bibinfo {author} {\bibfnamefont {Claudio}\ \bibnamefont {Chamon}}, \ and\
  \bibinfo {author} {\bibfnamefont {Titus}\ \bibnamefont {Neupert}},\
  }\bibfield  {title} {\enquote {\bibinfo {title} {Topological many-body scar
  states in dimensions one, two, and three},}\ }\href {\doibase
  10.1103/PhysRevResearch.1.033144} {\bibfield  {journal} {\bibinfo  {journal}
  {Phys. Rev. Research}\ }\textbf {\bibinfo {volume} {1}},\ \bibinfo {pages}
  {033144} (\bibinfo {year} {2019})}\BibitemShut {NoStop}%
\bibitem [{\citenamefont {Bull}\ \emph {et~al.}(2019)\citenamefont {Bull},
  \citenamefont {Martin},\ and\ \citenamefont {Papi\ifmmode~\acute{c}\else
  \'{c}\fi{}}}]{Bul-19}%
  \BibitemOpen
  \bibfield  {author} {\bibinfo {author} {\bibfnamefont {Kieran}\ \bibnamefont
  {Bull}}, \bibinfo {author} {\bibfnamefont {Ivar}\ \bibnamefont {Martin}}, \
  and\ \bibinfo {author} {\bibfnamefont {Z.}~\bibnamefont
  {Papi\ifmmode~\acute{c}\else \'{c}\fi{}}},\ }\bibfield  {title} {\enquote
  {\bibinfo {title} {Systematic construction of scarred many-body dynamics in
  1d lattice models},}\ }\href {\doibase 10.1103/PhysRevLett.123.030601}
  {\bibfield  {journal} {\bibinfo  {journal} {Phys. Rev. Lett.}\ }\textbf
  {\bibinfo {volume} {123}},\ \bibinfo {pages} {030601} (\bibinfo {year}
  {2019})}\BibitemShut {NoStop}%
\bibitem [{\citenamefont {Ho}\ \emph {et~al.}(2019)\citenamefont {Ho},
  \citenamefont {Choi}, \citenamefont {Pichler},\ and\ \citenamefont
  {Lukin}}]{Ho-19}%
  \BibitemOpen
  \bibfield  {author} {\bibinfo {author} {\bibfnamefont {Wen~Wei}\ \bibnamefont
  {Ho}}, \bibinfo {author} {\bibfnamefont {Soonwon}\ \bibnamefont {Choi}},
  \bibinfo {author} {\bibfnamefont {Hannes}\ \bibnamefont {Pichler}}, \ and\
  \bibinfo {author} {\bibfnamefont {Mikhail~D.}\ \bibnamefont {Lukin}},\
  }\bibfield  {title} {\enquote {\bibinfo {title} {Periodic orbits,
  entanglement, and quantum many-body scars in constrained models: Matrix
  product state approach},}\ }\href {\doibase 10.1103/PhysRevLett.122.040603}
  {\bibfield  {journal} {\bibinfo  {journal} {Phys. Rev. Lett.}\ }\textbf
  {\bibinfo {volume} {122}},\ \bibinfo {pages} {040603} (\bibinfo {year}
  {2019})}\BibitemShut {NoStop}%
\bibitem [{\citenamefont {Mark}\ \emph {et~al.}(2020)\citenamefont {Mark},
  \citenamefont {Lin},\ and\ \citenamefont {Motrunich}}]{Mar-20}%
  \BibitemOpen
  \bibfield  {author} {\bibinfo {author} {\bibfnamefont {Daniel~K.}\
  \bibnamefont {Mark}}, \bibinfo {author} {\bibfnamefont {Cheng-Ju}\
  \bibnamefont {Lin}}, \ and\ \bibinfo {author} {\bibfnamefont {Olexei~I.}\
  \bibnamefont {Motrunich}},\ }\bibfield  {title} {\enquote {\bibinfo {title}
  {Unified structure for exact towers of scar states in the
  affleck-kennedy-lieb-tasaki and other models},}\ }\href {\doibase
  10.1103/PhysRevB.101.195131} {\bibfield  {journal} {\bibinfo  {journal}
  {Phys. Rev. B}\ }\textbf {\bibinfo {volume} {101}},\ \bibinfo {pages}
  {195131} (\bibinfo {year} {2020})}\BibitemShut {NoStop}%
\bibitem [{\citenamefont {Shibata}\ \emph {et~al.}(2020)\citenamefont
  {Shibata}, \citenamefont {Yoshioka},\ and\ \citenamefont {Katsura}}]{Shi-20}%
  \BibitemOpen
  \bibfield  {author} {\bibinfo {author} {\bibfnamefont {Naoyuki}\ \bibnamefont
  {Shibata}}, \bibinfo {author} {\bibfnamefont {Nobuyuki}\ \bibnamefont
  {Yoshioka}}, \ and\ \bibinfo {author} {\bibfnamefont {Hosho}\ \bibnamefont
  {Katsura}},\ }\bibfield  {title} {\enquote {\bibinfo {title} {Onsager's scars
  in disordered spin chains},}\ }\href {\doibase
  10.1103/PhysRevLett.124.180604} {\bibfield  {journal} {\bibinfo  {journal}
  {Phys. Rev. Lett.}\ }\textbf {\bibinfo {volume} {124}},\ \bibinfo {pages}
  {180604} (\bibinfo {year} {2020})}\BibitemShut {NoStop}%
\bibitem [{\citenamefont {Moudgalya}\ \emph {et~al.}(2020)\citenamefont
  {Moudgalya}, \citenamefont {O'Brien}, \citenamefont {Bernevig}, \citenamefont
  {Fendley},\ and\ \citenamefont {Regnault}}]{Mou-20}%
  \BibitemOpen
  \bibfield  {author} {\bibinfo {author} {\bibfnamefont {Sanjay}\ \bibnamefont
  {Moudgalya}}, \bibinfo {author} {\bibfnamefont {Edward}\ \bibnamefont
  {O'Brien}}, \bibinfo {author} {\bibfnamefont {B~Andrei}\ \bibnamefont
  {Bernevig}}, \bibinfo {author} {\bibfnamefont {Paul}\ \bibnamefont
  {Fendley}}, \ and\ \bibinfo {author} {\bibfnamefont {Nicolas}\ \bibnamefont
  {Regnault}},\ }\bibfield  {title} {\enquote {\bibinfo {title} {Large classes
  of quantum scarred hamiltonians from matrix product states},}\ }\href@noop {}
  {\bibfield  {journal} {\bibinfo  {journal} {arXiv preprint arXiv:2002.11725}\
  } (\bibinfo {year} {2020})}\BibitemShut {NoStop}%
\bibitem [{\citenamefont {Michailidis}\ \emph {et~al.}(2020)\citenamefont
  {Michailidis}, \citenamefont {Turner}, \citenamefont {Papi{\'c}},
  \citenamefont {Abanin},\ and\ \citenamefont {Serbyn}}]{Mic-20}%
  \BibitemOpen
  \bibfield  {author} {\bibinfo {author} {\bibfnamefont {AA}~\bibnamefont
  {Michailidis}}, \bibinfo {author} {\bibfnamefont {CJ}~\bibnamefont {Turner}},
  \bibinfo {author} {\bibfnamefont {Z}~\bibnamefont {Papi{\'c}}}, \bibinfo
  {author} {\bibfnamefont {DA}~\bibnamefont {Abanin}}, \ and\ \bibinfo {author}
  {\bibfnamefont {Maksym}\ \bibnamefont {Serbyn}},\ }\bibfield  {title}
  {\enquote {\bibinfo {title} {Slow quantum thermalization and many-body
  revivals from mixed phase space},}\ }\href@noop {} {\bibfield  {journal}
  {\bibinfo  {journal} {Phys. Rev. X}\ }\textbf {\bibinfo {volume} {10}},\
  \bibinfo {pages} {011055} (\bibinfo {year} {2020})}\BibitemShut {NoStop}%
\bibitem [{\citenamefont {Iadecola}\ and\ \citenamefont
  {Schecter}(2020)}]{Iad-20}%
  \BibitemOpen
  \bibfield  {author} {\bibinfo {author} {\bibfnamefont {Thomas}\ \bibnamefont
  {Iadecola}}\ and\ \bibinfo {author} {\bibfnamefont {Michael}\ \bibnamefont
  {Schecter}},\ }\bibfield  {title} {\enquote {\bibinfo {title} {Quantum
  many-body scar states with emergent kinetic constraints and
  finite-entanglement revivals},}\ }\href {\doibase
  10.1103/PhysRevB.101.024306} {\bibfield  {journal} {\bibinfo  {journal}
  {Phys. Rev. B}\ }\textbf {\bibinfo {volume} {101}},\ \bibinfo {pages}
  {024306} (\bibinfo {year} {2020})}\BibitemShut {NoStop}%
\bibitem [{\citenamefont {Bull}\ \emph {et~al.}(2020)\citenamefont {Bull},
  \citenamefont {Desaules},\ and\ \citenamefont {Papi\ifmmode~\acute{c}\else
  \'{c}\fi{}}}]{Bul-20}%
  \BibitemOpen
  \bibfield  {author} {\bibinfo {author} {\bibfnamefont {Kieran}\ \bibnamefont
  {Bull}}, \bibinfo {author} {\bibfnamefont {Jean-Yves}\ \bibnamefont
  {Desaules}}, \ and\ \bibinfo {author} {\bibfnamefont {Zlatko}\ \bibnamefont
  {Papi\ifmmode~\acute{c}\else \'{c}\fi{}}},\ }\bibfield  {title} {\enquote
  {\bibinfo {title} {Quantum scars as embeddings of weakly broken lie algebra
  representations},}\ }\href {\doibase 10.1103/PhysRevB.101.165139} {\bibfield
  {journal} {\bibinfo  {journal} {Phys. Rev. B}\ }\textbf {\bibinfo {volume}
  {101}},\ \bibinfo {pages} {165139} (\bibinfo {year} {2020})}\BibitemShut
  {NoStop}%
\bibitem [{\citenamefont {Kuno}\ \emph {et~al.}(2020)\citenamefont {Kuno},
  \citenamefont {Mizoguchi},\ and\ \citenamefont {Hatsugai}}]{Kun-20}%
  \BibitemOpen
  \bibfield  {author} {\bibinfo {author} {\bibfnamefont {Yoshihito}\
  \bibnamefont {Kuno}}, \bibinfo {author} {\bibfnamefont {Tomonari}\
  \bibnamefont {Mizoguchi}}, \ and\ \bibinfo {author} {\bibfnamefont
  {Yasuhiro}\ \bibnamefont {Hatsugai}},\ }\bibfield  {title} {\enquote
  {\bibinfo {title} {Flat band quantum scar},}\ }\href {\doibase
  10.1103/PhysRevB.102.241115} {\bibfield  {journal} {\bibinfo  {journal}
  {Phys. Rev. B}\ }\textbf {\bibinfo {volume} {102}},\ \bibinfo {pages}
  {241115} (\bibinfo {year} {2020})}\BibitemShut {NoStop}%
\bibitem [{\citenamefont {McClarty}\ \emph {et~al.}(2020)\citenamefont
  {McClarty}, \citenamefont {Haque}, \citenamefont {Sen},\ and\ \citenamefont
  {Richter}}]{McC-20}%
  \BibitemOpen
  \bibfield  {author} {\bibinfo {author} {\bibfnamefont {Paul~A.}\ \bibnamefont
  {McClarty}}, \bibinfo {author} {\bibfnamefont {Masudul}\ \bibnamefont
  {Haque}}, \bibinfo {author} {\bibfnamefont {Arnab}\ \bibnamefont {Sen}}, \
  and\ \bibinfo {author} {\bibfnamefont {Johannes}\ \bibnamefont {Richter}},\
  }\bibfield  {title} {\enquote {\bibinfo {title} {Disorder-free localization
  and many-body quantum scars from magnetic frustration},}\ }\href {\doibase
  10.1103/PhysRevB.102.224303} {\bibfield  {journal} {\bibinfo  {journal}
  {Phys. Rev. B}\ }\textbf {\bibinfo {volume} {102}},\ \bibinfo {pages}
  {224303} (\bibinfo {year} {2020})}\BibitemShut {NoStop}%
\bibitem [{\citenamefont {Banerjee}\ and\ \citenamefont {Sen}(2020)}]{Ban-20}%
  \BibitemOpen
  \bibfield  {author} {\bibinfo {author} {\bibfnamefont {Debasish}\
  \bibnamefont {Banerjee}}\ and\ \bibinfo {author} {\bibfnamefont {Arnab}\
  \bibnamefont {Sen}},\ }\href@noop {} {\enquote {\bibinfo {title} {Quantum
  scars from zero modes in an abelian lattice gauge theory},}\ } (\bibinfo
  {year} {2020}),\ \Eprint {http://arxiv.org/abs/2012.08540} {arXiv:2012.08540
  [cond-mat.str-el]} \BibitemShut {NoStop}%
\bibitem [{\citenamefont {Dooley}\ and\ \citenamefont {Kells}(2020)}]{Doo-20b}%
  \BibitemOpen
  \bibfield  {author} {\bibinfo {author} {\bibfnamefont {Shane}\ \bibnamefont
  {Dooley}}\ and\ \bibinfo {author} {\bibfnamefont {Graham}\ \bibnamefont
  {Kells}},\ }\bibfield  {title} {\enquote {\bibinfo {title} {Enhancing the
  effect of quantum many-body scars on dynamics by minimizing the effective
  dimension},}\ }\href {\doibase 10.1103/PhysRevB.102.195114} {\bibfield
  {journal} {\bibinfo  {journal} {Phys. Rev. B}\ }\textbf {\bibinfo {volume}
  {102}},\ \bibinfo {pages} {195114} (\bibinfo {year} {2020})}\BibitemShut
  {NoStop}%
\bibitem [{\citenamefont {Dooley}\ \emph {et~al.}(2016)\citenamefont {Dooley},
  \citenamefont {Munro},\ and\ \citenamefont {Nemoto}}]{Doo-16b}%
  \BibitemOpen
  \bibfield  {author} {\bibinfo {author} {\bibfnamefont {Shane}\ \bibnamefont
  {Dooley}}, \bibinfo {author} {\bibfnamefont {William~J}\ \bibnamefont
  {Munro}}, \ and\ \bibinfo {author} {\bibfnamefont {Kae}\ \bibnamefont
  {Nemoto}},\ }\bibfield  {title} {\enquote {\bibinfo {title} {Quantum
  metrology including state preparation and readout times},}\ }\href@noop {}
  {\bibfield  {journal} {\bibinfo  {journal} {Physical Review A}\ }\textbf
  {\bibinfo {volume} {94}},\ \bibinfo {pages} {052320} (\bibinfo {year}
  {2016})}\BibitemShut {NoStop}%
\bibitem [{\citenamefont {Hayes}\ \emph {et~al.}(2018)\citenamefont {Hayes},
  \citenamefont {Dooley}, \citenamefont {Munro}, \citenamefont {Nemoto},\ and\
  \citenamefont {Dunningham}}]{Hay-18}%
  \BibitemOpen
  \bibfield  {author} {\bibinfo {author} {\bibfnamefont {Anthony~J}\
  \bibnamefont {Hayes}}, \bibinfo {author} {\bibfnamefont {Shane}\ \bibnamefont
  {Dooley}}, \bibinfo {author} {\bibfnamefont {William~J}\ \bibnamefont
  {Munro}}, \bibinfo {author} {\bibfnamefont {Kae}\ \bibnamefont {Nemoto}}, \
  and\ \bibinfo {author} {\bibfnamefont {Jacob}\ \bibnamefont {Dunningham}},\
  }\bibfield  {title} {\enquote {\bibinfo {title} {Making the most of time in
  quantum metrology: concurrent state preparation and sensing},}\ }\href
  {\doibase 10.1088/2058-9565/aac30b} {\bibfield  {journal} {\bibinfo
  {journal} {Quantum Science and Technology}\ }\textbf {\bibinfo {volume}
  {3}},\ \bibinfo {pages} {035007} (\bibinfo {year} {2018})}\BibitemShut
  {NoStop}%
\bibitem [{\citenamefont {Wineland}\ \emph {et~al.}(1994)\citenamefont
  {Wineland}, \citenamefont {Bollinger}, \citenamefont {Itano},\ and\
  \citenamefont {Heinzen}}]{Win-94}%
  \BibitemOpen
  \bibfield  {author} {\bibinfo {author} {\bibfnamefont {D.~J.}\ \bibnamefont
  {Wineland}}, \bibinfo {author} {\bibfnamefont {J.~J.}\ \bibnamefont
  {Bollinger}}, \bibinfo {author} {\bibfnamefont {W.~M.}\ \bibnamefont
  {Itano}}, \ and\ \bibinfo {author} {\bibfnamefont {D.~J.}\ \bibnamefont
  {Heinzen}},\ }\bibfield  {title} {\enquote {\bibinfo {title} {Squeezed atomic
  states and projection noise in spectroscopy},}\ }\href {\doibase
  10.1103/PhysRevA.50.67} {\bibfield  {journal} {\bibinfo  {journal} {Phys.
  Rev. A}\ }\textbf {\bibinfo {volume} {50}},\ \bibinfo {pages} {67--88}
  (\bibinfo {year} {1994})}\BibitemShut {NoStop}%
\bibitem [{\citenamefont {Deutsch}(1991)}]{Deu-91}%
  \BibitemOpen
  \bibfield  {author} {\bibinfo {author} {\bibfnamefont {J.~M.}\ \bibnamefont
  {Deutsch}},\ }\bibfield  {title} {\enquote {\bibinfo {title} {Quantum
  statistical mechanics in a closed system},}\ }\href {\doibase
  10.1103/PhysRevA.43.2046} {\bibfield  {journal} {\bibinfo  {journal} {Phys.
  Rev. A}\ }\textbf {\bibinfo {volume} {43}},\ \bibinfo {pages} {2046--2049}
  (\bibinfo {year} {1991})}\BibitemShut {NoStop}%
\bibitem [{\citenamefont {Srednicki}(1994)}]{Sre-94}%
  \BibitemOpen
  \bibfield  {author} {\bibinfo {author} {\bibfnamefont {Mark}\ \bibnamefont
  {Srednicki}},\ }\bibfield  {title} {\enquote {\bibinfo {title} {Chaos and
  quantum thermalization},}\ }\href {\doibase 10.1103/PhysRevE.50.888}
  {\bibfield  {journal} {\bibinfo  {journal} {Phys. Rev. E}\ }\textbf {\bibinfo
  {volume} {50}},\ \bibinfo {pages} {888--901} (\bibinfo {year}
  {1994})}\BibitemShut {NoStop}%
\bibitem [{\citenamefont {Rigol}\ \emph {et~al.}(2008)\citenamefont {Rigol},
  \citenamefont {Dunjko},\ and\ \citenamefont {Olshanii}}]{Rig-08}%
  \BibitemOpen
  \bibfield  {author} {\bibinfo {author} {\bibfnamefont {Marcos}\ \bibnamefont
  {Rigol}}, \bibinfo {author} {\bibfnamefont {Vanja}\ \bibnamefont {Dunjko}}, \
  and\ \bibinfo {author} {\bibfnamefont {Maxim}\ \bibnamefont {Olshanii}},\
  }\bibfield  {title} {\enquote {\bibinfo {title} {Thermalization and its
  mechanism for generic isolated quantum systems},}\ }\href@noop {} {\bibfield
  {journal} {\bibinfo  {journal} {Nature}\ }\textbf {\bibinfo {volume} {452}},\
  \bibinfo {pages} {854--858} (\bibinfo {year} {2008})}\BibitemShut {NoStop}%
\bibitem [{\citenamefont {D'Alessio}\ \emph {et~al.}(2016)\citenamefont
  {D'Alessio}, \citenamefont {Kafri}, \citenamefont {Polkovnikov},\ and\
  \citenamefont {Rigol}}]{DAl-16}%
  \BibitemOpen
  \bibfield  {author} {\bibinfo {author} {\bibfnamefont {Luca}\ \bibnamefont
  {D'Alessio}}, \bibinfo {author} {\bibfnamefont {Yariv}\ \bibnamefont
  {Kafri}}, \bibinfo {author} {\bibfnamefont {Anatoli}\ \bibnamefont
  {Polkovnikov}}, \ and\ \bibinfo {author} {\bibfnamefont {Marcos}\
  \bibnamefont {Rigol}},\ }\bibfield  {title} {\enquote {\bibinfo {title} {From
  quantum chaos and eigenstate thermalization to statistical mechanics and
  thermodynamics},}\ }\href@noop {} {\bibfield  {journal} {\bibinfo  {journal}
  {Advances in Physics}\ }\textbf {\bibinfo {volume} {65}},\ \bibinfo {pages}
  {239--362} (\bibinfo {year} {2016})}\BibitemShut {NoStop}%
\bibitem [{\citenamefont {Biroli}\ \emph {et~al.}(2010)\citenamefont {Biroli},
  \citenamefont {Kollath},\ and\ \citenamefont {L\"auchli}}]{Bir-10}%
  \BibitemOpen
  \bibfield  {author} {\bibinfo {author} {\bibfnamefont {Giulio}\ \bibnamefont
  {Biroli}}, \bibinfo {author} {\bibfnamefont {Corinna}\ \bibnamefont
  {Kollath}}, \ and\ \bibinfo {author} {\bibfnamefont {Andreas~M.}\
  \bibnamefont {L\"auchli}},\ }\bibfield  {title} {\enquote {\bibinfo {title}
  {Effect of rare fluctuations on the thermalization of isolated quantum
  systems},}\ }\href {\doibase 10.1103/PhysRevLett.105.250401} {\bibfield
  {journal} {\bibinfo  {journal} {Phys. Rev. Lett.}\ }\textbf {\bibinfo
  {volume} {105}},\ \bibinfo {pages} {250401} (\bibinfo {year}
  {2010})}\BibitemShut {NoStop}%
\bibitem [{\citenamefont {Shiraishi}\ and\ \citenamefont
  {Mori}(2017)}]{Shi-17}%
  \BibitemOpen
  \bibfield  {author} {\bibinfo {author} {\bibfnamefont {Naoto}\ \bibnamefont
  {Shiraishi}}\ and\ \bibinfo {author} {\bibfnamefont {Takashi}\ \bibnamefont
  {Mori}},\ }\bibfield  {title} {\enquote {\bibinfo {title} {Systematic
  construction of counterexamples to the eigenstate thermalization
  hypothesis},}\ }\href {\doibase 10.1103/PhysRevLett.119.030601} {\bibfield
  {journal} {\bibinfo  {journal} {Phys. Rev. Lett.}\ }\textbf {\bibinfo
  {volume} {119}},\ \bibinfo {pages} {030601} (\bibinfo {year}
  {2017})}\BibitemShut {NoStop}%
\bibitem [{\citenamefont {Dicke}(1954)}]{Dic-54}%
  \BibitemOpen
  \bibfield  {author} {\bibinfo {author} {\bibfnamefont {R.~H.}\ \bibnamefont
  {Dicke}},\ }\bibfield  {title} {\enquote {\bibinfo {title} {Coherence in
  spontaneous radiation processes},}\ }\href {\doibase 10.1103/PhysRev.93.99}
  {\bibfield  {journal} {\bibinfo  {journal} {Phys. Rev.}\ }\textbf {\bibinfo
  {volume} {93}},\ \bibinfo {pages} {99--110} (\bibinfo {year}
  {1954})}\BibitemShut {NoStop}%
\bibitem [{\citenamefont {Mark}\ and\ \citenamefont
  {Motrunich}(2020)}]{Mar-20b}%
  \BibitemOpen
  \bibfield  {author} {\bibinfo {author} {\bibfnamefont {Daniel~K.}\
  \bibnamefont {Mark}}\ and\ \bibinfo {author} {\bibfnamefont {Olexei~I.}\
  \bibnamefont {Motrunich}},\ }\bibfield  {title} {\enquote {\bibinfo {title}
  {$\ensuremath{\eta}$-pairing states as true scars in an extended hubbard
  model},}\ }\href {\doibase 10.1103/PhysRevB.102.075132} {\bibfield  {journal}
  {\bibinfo  {journal} {Phys. Rev. B}\ }\textbf {\bibinfo {volume} {102}},\
  \bibinfo {pages} {075132} (\bibinfo {year} {2020})}\BibitemShut {NoStop}%
\bibitem [{\citenamefont {Schecter}\ and\ \citenamefont
  {Iadecola}(2019)}]{Sch-19}%
  \BibitemOpen
  \bibfield  {author} {\bibinfo {author} {\bibfnamefont {Michael}\ \bibnamefont
  {Schecter}}\ and\ \bibinfo {author} {\bibfnamefont {Thomas}\ \bibnamefont
  {Iadecola}},\ }\bibfield  {title} {\enquote {\bibinfo {title} {Weak
  ergodicity breaking and quantum many-body scars in spin-1 ${XY}$ magnets},}\
  }\href {\doibase 10.1103/PhysRevLett.123.147201} {\bibfield  {journal}
  {\bibinfo  {journal} {Phys. Rev. Lett.}\ }\textbf {\bibinfo {volume} {123}},\
  \bibinfo {pages} {147201} (\bibinfo {year} {2019})}\BibitemShut {NoStop}%
\bibitem [{\citenamefont {Moreno}\ and\ \citenamefont
  {Parisio}(2018)}]{Mor-18b}%
  \BibitemOpen
  \bibfield  {author} {\bibinfo {author} {\bibfnamefont {M.~G.~M.}\
  \bibnamefont {Moreno}}\ and\ \bibinfo {author} {\bibfnamefont {Fernando}\
  \bibnamefont {Parisio}},\ }\href@noop {} {\enquote {\bibinfo {title} {All
  bipartitions of arbitrary {D}icke states},}\ } (\bibinfo {year} {2018}),\
  \Eprint {http://arxiv.org/abs/1801.00762} {arXiv:1801.00762 [quant-ph]}
  \BibitemShut {NoStop}%
\bibitem [{\citenamefont {Choi}\ \emph {et~al.}(2019)\citenamefont {Choi},
  \citenamefont {Turner}, \citenamefont {Pichler}, \citenamefont {Ho},
  \citenamefont {Michailidis}, \citenamefont {Papi\ifmmode~\acute{c}\else
  \'{c}\fi{}}, \citenamefont {Serbyn}, \citenamefont {Lukin},\ and\
  \citenamefont {Abanin}}]{Cho-19}%
  \BibitemOpen
  \bibfield  {author} {\bibinfo {author} {\bibfnamefont {Soonwon}\ \bibnamefont
  {Choi}}, \bibinfo {author} {\bibfnamefont {Christopher~J.}\ \bibnamefont
  {Turner}}, \bibinfo {author} {\bibfnamefont {Hannes}\ \bibnamefont
  {Pichler}}, \bibinfo {author} {\bibfnamefont {Wen~Wei}\ \bibnamefont {Ho}},
  \bibinfo {author} {\bibfnamefont {Alexios~A.}\ \bibnamefont {Michailidis}},
  \bibinfo {author} {\bibfnamefont {Zlatko}\ \bibnamefont
  {Papi\ifmmode~\acute{c}\else \'{c}\fi{}}}, \bibinfo {author} {\bibfnamefont
  {Maksym}\ \bibnamefont {Serbyn}}, \bibinfo {author} {\bibfnamefont
  {Mikhail~D.}\ \bibnamefont {Lukin}}, \ and\ \bibinfo {author} {\bibfnamefont
  {Dmitry~A.}\ \bibnamefont {Abanin}},\ }\bibfield  {title} {\enquote {\bibinfo
  {title} {Emergent {SU}(2) dynamics and perfect quantum many-body scars},}\
  }\href {\doibase 10.1103/PhysRevLett.122.220603} {\bibfield  {journal}
  {\bibinfo  {journal} {Phys. Rev. Lett.}\ }\textbf {\bibinfo {volume} {122}},\
  \bibinfo {pages} {220603} (\bibinfo {year} {2019})}\BibitemShut {NoStop}%
\bibitem [{\citenamefont {Lesanovsky}(2011)}]{Les-11}%
  \BibitemOpen
  \bibfield  {author} {\bibinfo {author} {\bibfnamefont {Igor}\ \bibnamefont
  {Lesanovsky}},\ }\bibfield  {title} {\enquote {\bibinfo {title} {Many-body
  spin interactions and the ground state of a dense {R}ydberg lattice gas},}\
  }\href {\doibase 10.1103/PhysRevLett.106.025301} {\bibfield  {journal}
  {\bibinfo  {journal} {Phys. Rev. Lett.}\ }\textbf {\bibinfo {volume} {106}},\
  \bibinfo {pages} {025301} (\bibinfo {year} {2011})}\BibitemShut {NoStop}%
\bibitem [{\citenamefont {Turner}\ \emph
  {et~al.}(2018{\natexlab{b}})\citenamefont {Turner}, \citenamefont
  {Michailidis}, \citenamefont {Abanin}, \citenamefont {Serbyn},\ and\
  \citenamefont {Papi\ifmmode~\acute{c}\else \'{c}\fi{}}}]{Tur-18b}%
  \BibitemOpen
  \bibfield  {author} {\bibinfo {author} {\bibfnamefont {C.~J.}\ \bibnamefont
  {Turner}}, \bibinfo {author} {\bibfnamefont {A.~A.}\ \bibnamefont
  {Michailidis}}, \bibinfo {author} {\bibfnamefont {D.~A.}\ \bibnamefont
  {Abanin}}, \bibinfo {author} {\bibfnamefont {M.}~\bibnamefont {Serbyn}}, \
  and\ \bibinfo {author} {\bibfnamefont {Z.}~\bibnamefont
  {Papi\ifmmode~\acute{c}\else \'{c}\fi{}}},\ }\bibfield  {title} {\enquote
  {\bibinfo {title} {Quantum scarred eigenstates in a {R}ydberg atom chain:
  Entanglement, breakdown of thermalization, and stability to perturbations},}\
  }\href {\doibase 10.1103/PhysRevB.98.155134} {\bibfield  {journal} {\bibinfo
  {journal} {Phys. Rev. B}\ }\textbf {\bibinfo {volume} {98}},\ \bibinfo
  {pages} {155134} (\bibinfo {year} {2018}{\natexlab{b}})}\BibitemShut
  {NoStop}%
\bibitem [{\citenamefont {Khemani}\ \emph {et~al.}(2019)\citenamefont
  {Khemani}, \citenamefont {Laumann},\ and\ \citenamefont {Chandran}}]{Khe-19}%
  \BibitemOpen
  \bibfield  {author} {\bibinfo {author} {\bibfnamefont {Vedika}\ \bibnamefont
  {Khemani}}, \bibinfo {author} {\bibfnamefont {Chris~R.}\ \bibnamefont
  {Laumann}}, \ and\ \bibinfo {author} {\bibfnamefont {Anushya}\ \bibnamefont
  {Chandran}},\ }\bibfield  {title} {\enquote {\bibinfo {title} {Signatures of
  integrability in the dynamics of {R}ydberg-blockaded chains},}\ }\href
  {\doibase 10.1103/PhysRevB.99.161101} {\bibfield  {journal} {\bibinfo
  {journal} {Phys. Rev. B}\ }\textbf {\bibinfo {volume} {99}},\ \bibinfo
  {pages} {161101} (\bibinfo {year} {2019})}\BibitemShut {NoStop}%
\bibitem [{\citenamefont {Lin}\ \emph {et~al.}(2020)\citenamefont {Lin},
  \citenamefont {Chandran},\ and\ \citenamefont {Motrunich}}]{Lin-20}%
  \BibitemOpen
  \bibfield  {author} {\bibinfo {author} {\bibfnamefont {Cheng-Ju}\
  \bibnamefont {Lin}}, \bibinfo {author} {\bibfnamefont {Anushya}\ \bibnamefont
  {Chandran}}, \ and\ \bibinfo {author} {\bibfnamefont {Olexei~I.}\
  \bibnamefont {Motrunich}},\ }\bibfield  {title} {\enquote {\bibinfo {title}
  {Slow thermalization of exact quantum many-body scar states under
  perturbations},}\ }\href {\doibase 10.1103/PhysRevResearch.2.033044}
  {\bibfield  {journal} {\bibinfo  {journal} {Phys. Rev. Research}\ }\textbf
  {\bibinfo {volume} {2}},\ \bibinfo {pages} {033044} (\bibinfo {year}
  {2020})}\BibitemShut {NoStop}%
\bibitem [{\citenamefont {Surace}\ \emph {et~al.}(2020)\citenamefont {Surace},
  \citenamefont {Votto}, \citenamefont {Lazo}, \citenamefont {Silva},
  \citenamefont {Dalmonte},\ and\ \citenamefont {Giudici}}]{Sur-20}%
  \BibitemOpen
  \bibfield  {author} {\bibinfo {author} {\bibfnamefont {Federica~Maria}\
  \bibnamefont {Surace}}, \bibinfo {author} {\bibfnamefont {Matteo}\
  \bibnamefont {Votto}}, \bibinfo {author} {\bibfnamefont {Eduardo~Gonzalez}\
  \bibnamefont {Lazo}}, \bibinfo {author} {\bibfnamefont {Alessandro}\
  \bibnamefont {Silva}}, \bibinfo {author} {\bibfnamefont {Marcello}\
  \bibnamefont {Dalmonte}}, \ and\ \bibinfo {author} {\bibfnamefont {Giuliano}\
  \bibnamefont {Giudici}},\ }\href@noop {} {\enquote {\bibinfo {title} {Exact
  many-body scars and their stability in constrained quantum chains},}\ }
  (\bibinfo {year} {2020}),\ \Eprint {http://arxiv.org/abs/2011.08218}
  {arXiv:2011.08218 [cond-mat.stat-mech]} \BibitemShut {NoStop}%
\bibitem [{\citenamefont {Viola}\ and\ \citenamefont {Lloyd}(1998)}]{Vio-98}%
  \BibitemOpen
  \bibfield  {author} {\bibinfo {author} {\bibfnamefont {Lorenza}\ \bibnamefont
  {Viola}}\ and\ \bibinfo {author} {\bibfnamefont {Seth}\ \bibnamefont
  {Lloyd}},\ }\bibfield  {title} {\enquote {\bibinfo {title} {Dynamical
  suppression of decoherence in two-state quantum systems},}\ }\href@noop {}
  {\bibfield  {journal} {\bibinfo  {journal} {Physical Review A}\ }\textbf
  {\bibinfo {volume} {58}},\ \bibinfo {pages} {2733} (\bibinfo {year}
  {1998})}\BibitemShut {NoStop}%
\bibitem [{\citenamefont {De~Lange}\ \emph {et~al.}(2010)\citenamefont
  {De~Lange}, \citenamefont {Wang}, \citenamefont {Riste}, \citenamefont
  {Dobrovitski},\ and\ \citenamefont {Hanson}}]{DeL-10}%
  \BibitemOpen
  \bibfield  {author} {\bibinfo {author} {\bibfnamefont {G}~\bibnamefont
  {De~Lange}}, \bibinfo {author} {\bibfnamefont {ZH}~\bibnamefont {Wang}},
  \bibinfo {author} {\bibfnamefont {D}~\bibnamefont {Riste}}, \bibinfo {author}
  {\bibfnamefont {VV}~\bibnamefont {Dobrovitski}}, \ and\ \bibinfo {author}
  {\bibfnamefont {R}~\bibnamefont {Hanson}},\ }\bibfield  {title} {\enquote
  {\bibinfo {title} {Universal dynamical decoupling of a single solid-state
  spin from a spin bath},}\ }\href@noop {} {\bibfield  {journal} {\bibinfo
  {journal} {Science}\ }\textbf {\bibinfo {volume} {330}},\ \bibinfo {pages}
  {60--63} (\bibinfo {year} {2010})}\BibitemShut {NoStop}%
\bibitem [{\citenamefont {Bluvstein}\ \emph {et~al.}(2021)\citenamefont
  {Bluvstein}, \citenamefont {Omran}, \citenamefont {Levine}, \citenamefont
  {Keesling}, \citenamefont {Semeghini}, \citenamefont {Ebadi}, \citenamefont
  {Wang}, \citenamefont {Michailidis}, \citenamefont {Maskara}, \citenamefont
  {Ho}, \citenamefont {Choi}, \citenamefont {Serbyn}, \citenamefont {Greiner},
  \citenamefont {Vuleti{\'c}},\ and\ \citenamefont {Lukin}}]{Blu-21}%
  \BibitemOpen
  \bibfield  {author} {\bibinfo {author} {\bibfnamefont {D.}~\bibnamefont
  {Bluvstein}}, \bibinfo {author} {\bibfnamefont {A.}~\bibnamefont {Omran}},
  \bibinfo {author} {\bibfnamefont {H.}~\bibnamefont {Levine}}, \bibinfo
  {author} {\bibfnamefont {A.}~\bibnamefont {Keesling}}, \bibinfo {author}
  {\bibfnamefont {G.}~\bibnamefont {Semeghini}}, \bibinfo {author}
  {\bibfnamefont {S.}~\bibnamefont {Ebadi}}, \bibinfo {author} {\bibfnamefont
  {T.~T.}\ \bibnamefont {Wang}}, \bibinfo {author} {\bibfnamefont {A.~A.}\
  \bibnamefont {Michailidis}}, \bibinfo {author} {\bibfnamefont
  {N.}~\bibnamefont {Maskara}}, \bibinfo {author} {\bibfnamefont {W.~W.}\
  \bibnamefont {Ho}}, \bibinfo {author} {\bibfnamefont {S.}~\bibnamefont
  {Choi}}, \bibinfo {author} {\bibfnamefont {M.}~\bibnamefont {Serbyn}},
  \bibinfo {author} {\bibfnamefont {M.}~\bibnamefont {Greiner}}, \bibinfo
  {author} {\bibfnamefont {V.}~\bibnamefont {Vuleti{\'c}}}, \ and\ \bibinfo
  {author} {\bibfnamefont {M.~D.}\ \bibnamefont {Lukin}},\ }\bibfield  {title}
  {\enquote {\bibinfo {title} {Controlling quantum many-body dynamics in driven
  {R}ydberg atom arrays},}\ }\href {\doibase 10.1126/science.abg2530}
  {\bibfield  {journal} {\bibinfo  {journal} {Science}\ } (\bibinfo {year}
  {2021}),\ 10.1126/science.abg2530}\BibitemShut {NoStop}%
\bibitem [{\citenamefont {Maskara}\ \emph {et~al.}(2021)\citenamefont
  {Maskara}, \citenamefont {Michailidis}, \citenamefont {Ho}, \citenamefont
  {Bluvstein}, \citenamefont {Choi}, \citenamefont {Lukin},\ and\ \citenamefont
  {Serbyn}}]{Mas-21}%
  \BibitemOpen
  \bibfield  {author} {\bibinfo {author} {\bibfnamefont {Nishad}\ \bibnamefont
  {Maskara}}, \bibinfo {author} {\bibfnamefont {Alexios~A}\ \bibnamefont
  {Michailidis}}, \bibinfo {author} {\bibfnamefont {Wen~Wei}\ \bibnamefont
  {Ho}}, \bibinfo {author} {\bibfnamefont {Dolev}\ \bibnamefont {Bluvstein}},
  \bibinfo {author} {\bibfnamefont {Soonwon}\ \bibnamefont {Choi}}, \bibinfo
  {author} {\bibfnamefont {Mikhail~D}\ \bibnamefont {Lukin}}, \ and\ \bibinfo
  {author} {\bibfnamefont {Maksym}\ \bibnamefont {Serbyn}},\ }\href@noop {}
  {\enquote {\bibinfo {title} {Discrete time-crystalline order enabled by
  quantum many-body scars: entanglement steering via periodic driving},}\ }
  (\bibinfo {year} {2021}),\ \Eprint {http://arxiv.org/abs/2102.13160}
  {arXiv:2102.13160 [quant-ph]} \BibitemShut {NoStop}%
\bibitem [{\citenamefont {Giovannetti}\ \emph {et~al.}(2004)\citenamefont
  {Giovannetti}, \citenamefont {Lloyd},\ and\ \citenamefont
  {Maccone}}]{Gio-04}%
  \BibitemOpen
  \bibfield  {author} {\bibinfo {author} {\bibfnamefont {Vittorio}\
  \bibnamefont {Giovannetti}}, \bibinfo {author} {\bibfnamefont {Seth}\
  \bibnamefont {Lloyd}}, \ and\ \bibinfo {author} {\bibfnamefont {Lorenzo}\
  \bibnamefont {Maccone}},\ }\bibfield  {title} {\enquote {\bibinfo {title}
  {Quantum-enhanced measurements: Beating the standard quantum limit},}\ }\href
  {\doibase 10.1126/science.1104149} {\bibfield  {journal} {\bibinfo  {journal}
  {Science}\ }\textbf {\bibinfo {volume} {306}},\ \bibinfo {pages} {1330--1336}
  (\bibinfo {year} {2004})}\BibitemShut {NoStop}%
\bibitem [{\citenamefont {Fujiwara}(2006)}]{Fuj-06}%
  \BibitemOpen
  \bibfield  {author} {\bibinfo {author} {\bibfnamefont {Akio}\ \bibnamefont
  {Fujiwara}},\ }\bibfield  {title} {\enquote {\bibinfo {title} {Strong
  consistency and asymptotic efficiency for adaptive quantum estimation
  problems},}\ }\href@noop {} {\bibfield  {journal} {\bibinfo  {journal}
  {Journal of Physics A: Mathematical and General}\ }\textbf {\bibinfo {volume}
  {39}},\ \bibinfo {pages} {12489} (\bibinfo {year} {2006})}\BibitemShut
  {NoStop}%
\bibitem [{\citenamefont {Okamoto}\ \emph {et~al.}(2012)\citenamefont
  {Okamoto}, \citenamefont {Iefuji}, \citenamefont {Oyama}, \citenamefont
  {Yamagata}, \citenamefont {Imai}, \citenamefont {Fujiwara},\ and\
  \citenamefont {Takeuchi}}]{Oka-12}%
  \BibitemOpen
  \bibfield  {author} {\bibinfo {author} {\bibfnamefont {Ryo}\ \bibnamefont
  {Okamoto}}, \bibinfo {author} {\bibfnamefont {Minako}\ \bibnamefont
  {Iefuji}}, \bibinfo {author} {\bibfnamefont {Satoshi}\ \bibnamefont {Oyama}},
  \bibinfo {author} {\bibfnamefont {Koichi}\ \bibnamefont {Yamagata}}, \bibinfo
  {author} {\bibfnamefont {Hiroshi}\ \bibnamefont {Imai}}, \bibinfo {author}
  {\bibfnamefont {Akio}\ \bibnamefont {Fujiwara}}, \ and\ \bibinfo {author}
  {\bibfnamefont {Shigeki}\ \bibnamefont {Takeuchi}},\ }\bibfield  {title}
  {\enquote {\bibinfo {title} {Experimental demonstration of adaptive quantum
  state estimation},}\ }\href {\doibase 10.1103/PhysRevLett.109.130404}
  {\bibfield  {journal} {\bibinfo  {journal} {Phys. Rev. Lett.}\ }\textbf
  {\bibinfo {volume} {109}},\ \bibinfo {pages} {130404} (\bibinfo {year}
  {2012})}\BibitemShut {NoStop}%
\bibitem [{\citenamefont {Yuan}\ and\ \citenamefont {Fung}(2015)}]{Yua-15}%
  \BibitemOpen
  \bibfield  {author} {\bibinfo {author} {\bibfnamefont {Haidong}\ \bibnamefont
  {Yuan}}\ and\ \bibinfo {author} {\bibfnamefont {Chi-Hang~Fred}\ \bibnamefont
  {Fung}},\ }\bibfield  {title} {\enquote {\bibinfo {title} {Optimal feedback
  scheme and universal time scaling for hamiltonian parameter estimation},}\
  }\href {\doibase 10.1103/PhysRevLett.115.110401} {\bibfield  {journal}
  {\bibinfo  {journal} {Phys. Rev. Lett.}\ }\textbf {\bibinfo {volume} {115}},\
  \bibinfo {pages} {110401} (\bibinfo {year} {2015})}\BibitemShut {NoStop}%
\bibitem [{\citenamefont {Wineland}\ \emph {et~al.}(1992)\citenamefont
  {Wineland}, \citenamefont {Bollinger}, \citenamefont {Itano}, \citenamefont
  {Moore},\ and\ \citenamefont {Heinzen}}]{Win-92}%
  \BibitemOpen
  \bibfield  {author} {\bibinfo {author} {\bibfnamefont {D.~J.}\ \bibnamefont
  {Wineland}}, \bibinfo {author} {\bibfnamefont {J.~J.}\ \bibnamefont
  {Bollinger}}, \bibinfo {author} {\bibfnamefont {W.~M.}\ \bibnamefont
  {Itano}}, \bibinfo {author} {\bibfnamefont {F.~L.}\ \bibnamefont {Moore}}, \
  and\ \bibinfo {author} {\bibfnamefont {D.~J.}\ \bibnamefont {Heinzen}},\
  }\bibfield  {title} {\enquote {\bibinfo {title} {Spin squeezing and reduced
  quantum noise in spectroscopy},}\ }\href {\doibase 10.1103/PhysRevA.46.R6797}
  {\bibfield  {journal} {\bibinfo  {journal} {Phys. Rev. A}\ }\textbf {\bibinfo
  {volume} {46}},\ \bibinfo {pages} {R6797--R6800} (\bibinfo {year}
  {1992})}\BibitemShut {NoStop}%
\bibitem [{\citenamefont {Kitagawa}\ and\ \citenamefont {Ueda}(1993)}]{Kit-93}%
  \BibitemOpen
  \bibfield  {author} {\bibinfo {author} {\bibfnamefont {Masahiro}\
  \bibnamefont {Kitagawa}}\ and\ \bibinfo {author} {\bibfnamefont {Masahito}\
  \bibnamefont {Ueda}},\ }\bibfield  {title} {\enquote {\bibinfo {title}
  {Squeezed spin states},}\ }\href@noop {} {\bibfield  {journal} {\bibinfo
  {journal} {Phys. Rev. A}\ }\textbf {\bibinfo {volume} {47}},\ \bibinfo
  {pages} {5138--5143} (\bibinfo {year} {1993})}\BibitemShut {NoStop}%
\bibitem [{\citenamefont {Ma}\ \emph {et~al.}(2011)\citenamefont {Ma},
  \citenamefont {Wang}, \citenamefont {Sun},\ and\ \citenamefont
  {Nori}}]{Ma-11}%
  \BibitemOpen
  \bibfield  {author} {\bibinfo {author} {\bibfnamefont {Jian}\ \bibnamefont
  {Ma}}, \bibinfo {author} {\bibfnamefont {Xiaoguang}\ \bibnamefont {Wang}},
  \bibinfo {author} {\bibfnamefont {C.P.}\ \bibnamefont {Sun}}, \ and\ \bibinfo
  {author} {\bibfnamefont {Franco}\ \bibnamefont {Nori}},\ }\bibfield  {title}
  {\enquote {\bibinfo {title} {Quantum spin squeezing},}\ }\href {\doibase
  http://dx.doi.org/10.1016/j.physrep.2011.08.003} {\bibfield  {journal}
  {\bibinfo  {journal} {Physics Reports}\ }\textbf {\bibinfo {volume} {509}},\
  \bibinfo {pages} {89 -- 165} (\bibinfo {year} {2011})}\BibitemShut {NoStop}%
\bibitem [{\citenamefont {Huelga}\ \emph {et~al.}(1997)\citenamefont {Huelga},
  \citenamefont {Macchiavello}, \citenamefont {Pellizzari}, \citenamefont
  {Ekert}, \citenamefont {Plenio},\ and\ \citenamefont {Cirac}}]{Hue-97}%
  \BibitemOpen
  \bibfield  {author} {\bibinfo {author} {\bibfnamefont {S.~F.}\ \bibnamefont
  {Huelga}}, \bibinfo {author} {\bibfnamefont {C.}~\bibnamefont
  {Macchiavello}}, \bibinfo {author} {\bibfnamefont {T.}~\bibnamefont
  {Pellizzari}}, \bibinfo {author} {\bibfnamefont {A.~K.}\ \bibnamefont
  {Ekert}}, \bibinfo {author} {\bibfnamefont {M.~B.}\ \bibnamefont {Plenio}}, \
  and\ \bibinfo {author} {\bibfnamefont {J.~I.}\ \bibnamefont {Cirac}},\
  }\bibfield  {title} {\enquote {\bibinfo {title} {Improvement of frequency
  standards with quantum entanglement},}\ }\href {\doibase
  10.1103/PhysRevLett.79.3865} {\bibfield  {journal} {\bibinfo  {journal}
  {Phys. Rev. Lett.}\ }\textbf {\bibinfo {volume} {79}},\ \bibinfo {pages}
  {3865--3868} (\bibinfo {year} {1997})}\BibitemShut {NoStop}%
\bibitem [{\citenamefont {Escher}\ \emph {et~al.}(2011)\citenamefont {Escher},
  \citenamefont {de~Matos~Filho},\ and\ \citenamefont {Davidovich}}]{Esc-11}%
  \BibitemOpen
  \bibfield  {author} {\bibinfo {author} {\bibfnamefont {B.~M.}\ \bibnamefont
  {Escher}}, \bibinfo {author} {\bibfnamefont {R.~L.}\ \bibnamefont
  {de~Matos~Filho}}, \ and\ \bibinfo {author} {\bibfnamefont {L.}~\bibnamefont
  {Davidovich}},\ }\bibfield  {title} {\enquote {\bibinfo {title} {General
  framework for estimating the ultimate precision limit in noisy
  quantum-enhanced metrology},}\ }\href {\doibase 10.1038/nphys1958} {\bibfield
   {journal} {\bibinfo  {journal} {Nat Phys}\ }\textbf {\bibinfo {volume}
  {7}},\ \bibinfo {pages} {406--411} (\bibinfo {year} {2011})}\BibitemShut
  {NoStop}%
\bibitem [{\citenamefont {Demkowicz-Dobrzanski}\ \emph
  {et~al.}(2012)\citenamefont {Demkowicz-Dobrzanski}, \citenamefont
  {Kolodynski},\ and\ \citenamefont {Guta}}]{Dem-12}%
  \BibitemOpen
  \bibfield  {author} {\bibinfo {author} {\bibfnamefont {Rafal}\ \bibnamefont
  {Demkowicz-Dobrzanski}}, \bibinfo {author} {\bibfnamefont {Jan}\ \bibnamefont
  {Kolodynski}}, \ and\ \bibinfo {author} {\bibfnamefont {Madalin}\
  \bibnamefont {Guta}},\ }\bibfield  {title} {\enquote {\bibinfo {title} {The
  elusive {H}eisenberg limit in quantum-enhanced metrology},}\ }\href
  {http://dx.doi.org/10.1038/ncomms2067} {\bibfield  {journal} {\bibinfo
  {journal} {Nat Commun}\ }\textbf {\bibinfo {volume} {3}},\ \bibinfo {pages}
  {1063} (\bibinfo {year} {2012})}\BibitemShut {NoStop}%
\bibitem [{\citenamefont {Chaves}\ \emph {et~al.}(2013)\citenamefont {Chaves},
  \citenamefont {Brask}, \citenamefont {Markiewicz}, \citenamefont
  {Ko\l{}ody\ifmmode~\acute{n}\else \'{n}\fi{}ski},\ and\ \citenamefont
  {Ac\'{\i}n}}]{Cha-13}%
  \BibitemOpen
  \bibfield  {author} {\bibinfo {author} {\bibfnamefont {R.}~\bibnamefont
  {Chaves}}, \bibinfo {author} {\bibfnamefont {J.~B.}\ \bibnamefont {Brask}},
  \bibinfo {author} {\bibfnamefont {M.}~\bibnamefont {Markiewicz}}, \bibinfo
  {author} {\bibfnamefont {J.}~\bibnamefont {Ko\l{}ody\ifmmode~\acute{n}\else
  \'{n}\fi{}ski}}, \ and\ \bibinfo {author} {\bibfnamefont {A.}~\bibnamefont
  {Ac\'{\i}n}},\ }\bibfield  {title} {\enquote {\bibinfo {title} {Noisy
  metrology beyond the standard quantum limit},}\ }\href {\doibase
  10.1103/PhysRevLett.111.120401} {\bibfield  {journal} {\bibinfo  {journal}
  {Phys. Rev. Lett.}\ }\textbf {\bibinfo {volume} {111}},\ \bibinfo {pages}
  {120401} (\bibinfo {year} {2013})}\BibitemShut {NoStop}%
\end{thebibliography}%


\onecolumngrid
\appendix

\section{The estimation error in Eq. \ref{eq:error_fit} is independent of $\omega$} \label{app:omega_independence}

The expression for the estimation error in Eq. \ref{eq:error_fit} is independent of the target parameter $\omega$. This $\omega$-independence is a consequence of two features of our scheme: (i) the symmetry $[\hat{H}_0, \hat{H}_\text{int}] = 0$, where $\hat{H}_0 =  \frac{\omega}{2}\sum_n \hat{S}_n^z$ and $\hat{H}_\text{int} = \sum_{n.n'}(e^{i\phi}\hat{S}_n^+ \hat{S}_n^- + \text{h.c.})$, and (ii) the optimisation of the error over the variable $\theta$ in our measurement observable $\hat{\mathcal{O}}_\theta = e^{-i\theta}\sum_n (\hat{S}_n^+)^2 + \text{h.c.}$. To see this, consider the measurement observable in the Heisenberg picture, $\hat{\mathcal{O}}_\theta (t) = e^{it\hat{H}}\hat{\mathcal{O}}_\theta e^{-it\hat{H}}$, where $\hat{H} = \hat{H}_0 + \hat{H}_\text{int}$. Using the property $[\hat{H}_0, \hat{H}_\text{int}] = 0$ we can rewrite this as: \begin{equation} \hat{\mathcal{O}}_\theta (t) = e^{it\hat{H}_\text{int}} e^{it\hat{H}_0} \hat{\mathcal{O}}_\theta e^{-it\hat{H}_0}e^{-it\hat{H}_\text{int}} = e^{it\hat{H}_\text{int}} \hat{\mathcal{O}}_{\theta - 2t\omega} e^{-it\hat{H}_\text{int}} , \nonumber\end{equation} where, in the last equality, the effect of the $\omega$-dependent $\hat{H}_0$ term has been absorbed into the measurement observable with $\theta \to \theta - 2t\omega$. Since the variable $\theta - 2\omega$ is optimised in the final measurement, the estimation error will be completely independent of $\omega$. Of course, the optimised measurement observable will then depend on the unknown parameter $\omega$. However, this is not a major problem in the limit of many repetitions $T/t \gg 1$ of the measurement: adaptive measurement schemes can begin with a non-optimal random value of $\theta$ and push towards the optimal value as the estimate of $\omega$ improves \cite{Fuj-06,Oka-12}.

If we break property (i), for example by adding the transverse field $\hat{H}_\Omega = \frac{\Omega}{2} \sum_n [e^{i\eta}(\hat{S}_n^+)^2 + \text{h.c.}]$ to the non-interacting part of the Hamiltonian, then the estimation error will no longer be independent of $\omega$. Indeed, if $\Omega \gg \omega$ the sensitivity to $\omega$ will be severely degraded. As mentioned in the main text, one approach to this is to measure the total field strength $\sqrt{\omega^2 + \Omega^2}$ instead of just the $z$-direction field $\omega$. However, if it is important to measure $\omega$ rather than $\sqrt{\omega^2 + \Omega^2}$, another approach is to apply a large known field $\omega_\text{known}$ in the $z$-direction, so that $\omega + \omega_\text{known} \gg \Omega$. After estimating the total value $\omega + \omega_\text{known}$ the known value can be subtracted to give an estimate of the unknown field $\omega$. Yet another approach would be to apply time-dependent quantum controls: it has been shown that this can restore the standard quantum limit for estimating $\omega$, even when $\omega \ll \Omega$ \cite{Yua-15}.

\section{The Dicke states are scars of the spin-1 DMI Hamiltonian, $H(\phi=\pm\pi/2)$}\label{app:Dicke_scars}

Here we prove that the Dicke states $\{\ket{\Psi(s)}\}_{s=0}^N$ are eigenstates of the spin-1 DMI Hamiltonian $\hat{H} = \hat{H}_0 + \sum_{n,n'}\lambda_{n,n'} \hat{h}_{n,n'}$, where $\hat{H}_0 = \frac{\omega}{2}\sum_{n=0}^{N-1}\hat{S}_n^z$ and $\hat{h}_{n,n'} = i\hat{S}_n^+\hat{S}_{n'}^- - \hat{S}_n^-\hat{S}_{n'}^+$. (We have assumed $\phi = \pi/2$ but the proof is identical for $\phi = -\pi/2$.) We note that this was already proved in the Appendix of Ref. \cite{Mar-20b}, and that the proof here is based on the one given in Ref. \cite{Sch-19} for a similar spin-1 model.

We know that the symmetric Dicke states are eigenstates of the non-interacting Hamiltonian $\hat{H}_0$, since this is one of the defining properties of Dicke states. To show that they are also eigenstates of the interacting part we will show that $\hat{h}_{n,n'}\ket{\Psi(s)} = 0$ for all $n,n'$ (following Ref. \cite{Sch-19}). First rewrite the Dicke state as the superposition: \begin{equation} \ket{\Psi(s)} = \binom{N}{s}^{-\frac{1}{2}}\sum_{\text{perms}} \ket{m=+1}^{\otimes s} \ket{m=-1}^{\otimes (N-s)} , \label{eq:Dicke_superposition} \end{equation} where the sum is over all permutations of $s$ spins in the state $\ket{m=+1}$ and the remaining $N-s$ spins in the state $\ket{m=-1}$. Any states in the superposition with $m_n = m_{n'} = 1$ or $m_n = m_{n'} = - 1$ are annihilated by $\hat{h}_{n,n'}$, since $\hat{S}_n^\pm \hat{S}_{n'}^\mp \ket{m_n=1}\ket{m_{n'} =1} = \hat{S}_n^\pm \hat{S}_{n'}^\mp \ket{m_n=-1}\ket{m_{n'} =-1} = 0$. Suppose that there is a term in the superposition Eq. \ref{eq:Dicke_superposition} of the form $\ket{A}\ket{-1_n}\ket{B}\ket{+1_{n'}}\ket{C}$ (i.e., with $m_n = -1$, $m_{n'} = +1$, and with $A$, $B$, and $C$ representing strings of $\pm 1$'s for the other spins). Then the term $\ket{A}\ket{+1_n}\ket{B}\ket{-1_{n'}}\ket{C}$ must also be present in the superposition, since the superposition includes \emph{all} permutations of the spins. But we have: \begin{eqnarray} \hat{h}_{n,n'}(\ket{A}\ket{-1_n}\ket{B}\ket{+1_{n'}}\ket{C} + \ket{A}\ket{+1_n}\ket{B}\ket{-1_{n'}}\ket{C}) &=& i \ket{A}\ket{0_n}\ket{B}\ket{0_{n'}}\ket{C} - i \ket{A}\ket{0_n}\ket{B}\ket{0_{n'}}\ket{C} \nonumber \\ &=& 0 . \end{eqnarray} Since this this covers all possiblities $m_n, m_{n'} \in \{ \pm 1 \}$, all states in the superposition Eq. \ref{eq:Dicke_superposition} are either annihilated or cancelled out, and we have $h_{n,n'} \ket{\Psi(s)} = 0$.

Since $\hat{H}_\text{int}\ket{\Psi(s)} = 0$ for all $s$, any superposition of the states $\{ \ket{\Psi(s)} \}_{s=0}^N$ will also be annihilated by $\hat{H}_\text{int}$. In particular, let $\hat{U}$ be any unitary transformation that leaves the symmetric Dicke subspace invariant, i.e., $\hat{U}$ has the property that $[\hat{U}, \sum_s \ket{\Psi(s)}\bra{\Psi(s)}] = 0$. Then $\hat{U}\ket{\Psi(s)}$ is still in the symmetric Dicke subspace and so we have $\hat{H}_\text{int}\hat{U}\ket{\Psi(s)} = 0$. The rotated states $\hat{U}\ket{\Psi(s)}$ will no longer by eigenstates of the non-interacting Hamiltonian $\hat{H}_0 = \frac{\omega}{2}\sum_{n=0}^{N-1}\hat{S}_n^z$, but they will be eigenstates of $\hat{U}\hat{H}_0 \hat{U}^\dagger$. This means that $\{\hat{U}\ket{\Psi(s)}\}_{s=0}^N$ are eigenstates of the Hamiltonian $\hat{H}_{U} = \hat{U}\hat{H}_0\hat{U}^\dagger + \hat{H}_\text{int}$. Towards the end of Sec. \ref{sec:spin_1_QMBS}, this was used to show that the scar states can be present even when the magnetisation symmetry of the Hamiltonian is broken by $\hat{U}$.





\section{Noise suppression with periodic control in spin-1 example}\label{app:noise}

In section \ref{sec:noise} we showed numerically that periodic application of the $\pi$-pulse $\hat{V}_{\pi} \equiv e^{i\pi\hat{S}_0^x}e^{i\pi\hat{S}_1^y}e^{i\pi\hat{S}_2^x}e^{i\pi\hat{S}_3^y}\hdots$ suppresses the unwanted XX-interaction and enhances estimation of an alternating signal. Here we show that in the limit of high-frequency control the XX-interaction, as well as noise due to inhomogeneous local magnetic fields, are completely eliminated leaving only the DMI to which the sensing is already robust.

The Hamiltonian is $\hat{H}(t) = \hat{H}_0(t) + \hat{H}_\Delta + \hat{H}_\text{XX} + \hat{H}_\text{DMI}$, where $\hat{H}_0(t) = \frac{\omega}{2} \sin(\pi t/\tau) \sum_n \hat{S}_n^z$ is our alternating signal, $\hat{H}_\Delta = \sum_n \Delta_n \hat{S}_n^z$ is the local magnetic field noise, $\hat{H}_\text{XX} = \lambda \cos\phi \sum_n (\hat{S}_n^x \hat{S}_{n+1}^x + \hat{S}_n^y \hat{S}_{n+1}^y)$ is the $XX$-interaction, and $\hat{H}_\text{DMI} = \lambda\sin\phi\sum_n (\hat{S}_n^x \hat{S}_{n+1}^y - \hat{S}_n^y \hat{S}_{n+1}^x)$ is the DMI. The corresponding unitary time-evolution operator between times $t_j$ and $t_{j+1}$ is $\hat{U}(t_{j+1}, t_j) = \mathcal{T}\exp[-i\int_{t_j}^{t_{j+1}} dt \hat{H}(t)]$. If we apply the $\pi$-pulse $\hat{V}_{\pi}$ periodically, at times $t_j = j\tau$, where $j=1,2,...,m$, then the total time-evolution operator, including the controls is: \begin{equation} \hat{U}_\text{ctrl}(m\tau) = \hat{V}_{\pi} \hat{U}(t_m,t_{m-1}) \hat{V}_{\pi} \hat{U}(t_{m-1},t_{m-2}) \hat{V}_{\pi} \hat{U}(t_{m-2}, t_{m-3}) \hdots \hat{V}_{\pi} \hat{U}(t_1, t_0) . \end{equation} If the inter-pulse period $\tau$ is sufficiently short, we can approximate: \begin{eqnarray} \hat{U}(t_{j},t_{j-1}) &\approx& e^{-i\tau \hat{H}_\Delta} e^{-i\tau\hat{H}_{XX}} e^{-i\tau\hat{H}_\text{DMI}} e^{-i\int_{t_{j-1}}^{t_j} dt \hat{H}_0(t)} , \\ \hat{V}_{\pi} \hat{U}(t_{j+1},t_{j}) \hat{V}_{\pi} &\approx& e^{+i\tau \hat{H}_\Delta} e^{+i\tau\hat{H}_{XX}} e^{-i\tau\hat{H}_\text{DMI}} e^{+i\int_{t_{j}}^{t_{j+1}} dt \hat{H}_0(t)} , \end{eqnarray} where the exponentials commute with each other up to $\mathcal{O}(\tau)$. The total evolution operator is then approximately: \begin{eqnarray} \hat{U}_\text{ctrl}(m\tau) \approx \exp\left[ -im\tau \hat{H}_\text{DMI} + i \sum_{j=1}^{m} (-1)^j \int_{t_{j-1}}^{t_j} dt \hat{H}_0(t) \right] = \exp\left[ -im\tau \hat{H}_\text{DMI} + i m\tau (\omega / \pi) \sum_n \hat{S}_n^z  \right] . \end{eqnarray} We see that the noise terms are completely suppressed, and the effective dynamics are given by the DMI and a static signal. The effective static signal frequency is suppressed by a factor of $2/\pi$ compared to the $\phi=\pi/2$ ideal case in section \ref{sec:spin_1_sensing}. The estimation error is therefore larger by the same factor, and we have $\delta\omega = 2 \delta\omega_\text{SQL}/\pi = 2/\sqrt{\pi^2 Nt T}$.

\section{Squeezing-enhanced sensing in the spin-1 DMI model}\label{app:squeezing}

It is well known that for non-interacting spin systems, the sensing performance can sometimes be enhanced by using an entangled or spin squeezed initial state instead of a separable initial state \cite{Gio-04}. Then the error can be decreased to the Heisenberg limit $\delta\omega_\text{HL} \propto 1 / (N\sqrt{tT})$, a $\sqrt{N}$ scaling enhancement compared to the standard quantum limit $\delta\omega_\text{SQL} \propto 1/\sqrt{NtT}$.

In principle, such a quantum enhancement is also possible here, for an interacting system with many-body scars. This is seen most clearly in our spin-1 interacting model. In Sec. \ref{sec:spin_1_sensing} we assumed that the probe was prepared in the separable initial state $\ket{\boldsymbol{+}}$ (defined in Eq. \ref{eq:psi_plus}), leading to an estimation error $\delta\omega = \delta\omega_\text{SQL} = 1/\sqrt{NtT}$, even in the presence of strong interactions (if $\phi = \pm \pi/2$). More generally, however, if the probe is prepared in a squeezed initial state the estimation error can be decreased by a factor $\xi < 1$, to $\delta\omega = \xi \delta\omega_\text{SQL}$. This enhancement factor is the squeezing parameter first defined by Wineland \emph{et al.} \cite{Win-92, Win-94}: \begin{equation} \xi = \frac{\sqrt{N}}{|\langle  \vec{\hat{J}} \rangle|} \min_{\vec{r}_\perp}\Delta (\vec{r}_\perp \cdot \vec{\hat{J}}) , \end{equation} where the minimisation is over all unit vectors $\vec{r}_\perp$ that are perpendicular to the mean spin direction $\langle  \vec{\hat{J}} \rangle$ of the initial state, and $\Delta (\vec{r}_\perp \cdot \vec{\hat{J}})$ is the standard deviation of the operator $\vec{r}_\perp \cdot \vec{\hat{J}}$. We note that the operators $\vec{\hat{J}} = (\hat{J}^x, \hat{J}^y, \hat{J}^z)$ here refer to the spin-1/2 system ``embedded'' in the spin-1 system, as explained at the beginning of Sec. \ref{sec:spin_1_QMBS}.


To generate the squeezed initial state in our spin-1 model we could, for example, prepare the two-axis twisted state $\hat{S}(\chi) \ket{\boldsymbol{+}}$, where $\hat{S}(\chi) = \exp [ i \chi (\hat{J}^y \hat{J}^z + \hat{J}^z \hat{J}^y) ]$ \cite{Kit-93, Ma-11}. For $N \gg 1$, the optimal squeezing strength $\chi = \chi_\text{opt}$ results in a squeezing parameter $\xi \sim 1/\sqrt{N}$ \cite{Kit-93, Ma-11}, so that the error is at the Heisenberg limit $\delta\omega = \xi\delta\omega_\text{SQL} \sim 1/(N\sqrt{tT})$ \cite{Win-92}.

Although sensing using entangled or squeezed states can be very fragile against some types of noise \cite{Hue-97, Esc-11, Dem-12}, it is known that significant gains can still be achieved in certain noisy scenarios \cite{Cha-13, Tan-15}.

\end{document}